\documentclass[reprint,amsmath,amssymb,aps,pre]{revtex4-2}
\usepackage[caption=false,font=footnotesize,indention=0cm]{subfig}
\usepackage{bm}
\usepackage{tabularx}
\usepackage{graphicx}
\usepackage{multirow}
\usepackage{hyperref}

\def\R{\mathbb{R}}
\def\G{\mathcal{G}}
\def\V{\mathcal{V}}
\def\E{\mathcal{E}}
\def\Gl{\mathcal{G}^{(l)}}
\def\Vl{\mathcal{V}^{(l)}}
\def\Vk{\mathcal{V}^{(k)}}
\def\El{\mathcal{E}^{(l)}}
\def\xl{x^{(l)}}
\def\xk{x^{(k)}}
\def\wl{w^{(l)}}
\def\Al{\bm{A}^{(l)}}
\def\Aintra{\bm{A}_{\mathrm{intra}}}
\def\Ainter{\bm{A}_{\mathrm{inter}}}
\def\A1{\bm{A}^{(1)}}
\def\AL{\bm{A}^{(L)}}
\def\x1{x^{(1)}}
\def\lmin{\lambda_{\mathrm{min}}}
\def\lmax{\lambda_{\mathrm{max}}}

\begin{document}

\title{Fast computation of matrix function-based centrality measures for layer-coupled multiplex networks}

\author{Kai Bergermann}
\email[]{kai.bergermann@math.tu-chemnitz.de}
\author{Martin Stoll}
\email[]{martin.stoll@math.tu-chemnitz.de}
\affiliation{Department of Mathematics, Technische Universit\"at Chemnitz, 09107 Chemnitz, Germany}
\date{\today}

\begin{abstract}
Centrality measures identify and rank the most influential entities of complex networks.
In this paper, we generalize matrix function-based centrality measures, which have been studied extensively for single-layer and temporal networks in recent years to layer-coupled multiplex networks.
The layers of these networks can reflect different relationships and interactions between entities or changing interactions over time.
We use the supra-adjacency matrix as network representation, which has already been used to generalize eigenvector centrality to temporal and multiplex networks.
With a suitable choice of edge weights, the definition of single-layer matrix function-based centrality measures in terms of walks on networks carries over naturally to the multilayer case.
In contrast to other walk-based centralities, matrix function-based centralities are parameterized measures, which have been shown to interpolate between (local) degree and (global) eigenvector centrality in the single-layer case.
As the explicit evaluation of the involved matrix function expressions becomes infeasible for medium to large-scale networks, we present highly efficient approximation techniques from numerical linear algebra, which rely on Krylov subspace methods, Gauss quadrature, and stochastic trace estimation.
We present extensive numerical studies on synthetic and real-world multiplex transportation, communication, and collaboration networks.
The comparison with established multilayer centrality measures shows that our framework produces meaningful rankings of nodes, layers, and node-layer pairs.
Furthermore, our experiments corroborate the linear computational complexity of the employed numerical methods in terms of the network size that is theoretically indicated under the assumption of sparsity in the supra-adjacency matrix.
This excellent scalability allows the efficient treatment of large-scale networks with the number of node-layer pairs of order $10^7$ or higher.
\end{abstract}

\maketitle

\section{Introduction}\label{sec:Introduction}

The study of complex networks has been a thriving interdisciplinary endeavor for many decades and some of the most impactful results found their way into our daily life \cite{milgram1967small,watts1998collective,barabasi1999emergence,brin1998anatomy,page1999pagerank}.
Applications of network science range from biology, chemistry, and physics over engineering and economics to the social sciences, cf.~\cite{estrada2012structure,kivela2014multilayer} and the references therein.
In recent years, much effort has been devoted to the generalization of established network-based methods to the case of multilayer structures, cf.~e.g., \cite{mucha2010community,kivela2014multilayer,boccaletti2014structure,sole2016random,taylor2017eigenvector,taylor2019supracentrality,taylor2021tunable,bergermann2021semi}.
These allow entities to interact in several different ways, reflect different types of relationships or changing interactions over time leading to ever more realistic models of highly complex phenomena.

The problem of identifying and ranking the most central nodes, i.e., entities of a network has a long history.
The variety of established centrality measures includes degree centrality, betweenness centrality \cite{freeman1977set}, closeness centrality \cite{freeman1978centrality}, eigenvector centrality \cite{bonacich1987power}, and variants of eigenvector centrality, which were developed in the context of the early internet \cite{brin1998anatomy,page1999pagerank,kleinberg1999authoritative}.
To date, the study of centrality measures has become a very active field of research and some recent works include \cite{gleich2015pagerank,aprahamian2016matching,fenu2017block,taylor2017eigenvector,wang2017identifying,chen2017dynamic,arrigo2017sparse,tudisco2018node,arrigo2018non,arrigo2018exponential,taylor2019supracentrality,benson2019three,wu2019tensor,benzi2020matrix,taylor2021tunable,al2021block,arrigo2021dynamic}.

This paper addresses the class of matrix function-based centrality and communicability measures, which have been studied intensively for single-layer networks \cite{katz1953new,estrada2000characterization,estrada2005subgraph,estrada2008communicability,estrada2010network,benzi2013ranking,benzi2013total,benzi2020matrix} and dynamic networks \cite{grindrod2011communicability,grindrod2013matrix,grindrod2014dynamical,chen2017dynamic,fenu2017block,arrigo2017sparse,al2021block,arrigo2021dynamic}.
These parameterized measures can be tuned to emphasize subgraphs of different sizes and have been shown to interpolate between (local) degree and (global) eigenvector centrality in the single-layer case \cite{benzi2015limiting}.

The main contribution of this paper is twofold: we generalize matrix function-based centrality measures to a general class of multilayer networks and we present numerical methods for the fast computation of the involved matrix function expressions.
The latter relies on highly efficient techniques from numerical linear algebra, which effectively scale to large-scale networks.

The multilayer networks considered in this paper are (node-aligned) layer-coupled multiplex networks in which the layers can represent different relationships and interactions or changing interactions between the same entities over time.
We differentiate between intra-layer edges connecting nodes from the same layer and inter-layer edges connecting nodes belonging to different layers and allow all edges to be directed or undirected.
In the case of directed networks we differentiate between each entity's role as broadcaster and receiver.
While it is not required from a theoretical perspective, we restrict inter-layer edges to only connect instances of the same physical node, i.e., copies of the same node in different layers.
Furthermore, we fix the edge weights between pairs of layers for all inter-layer edges between them.
This choice of multilayer network structure is particularly well-suited to create meaningful multilayer networks from multiple single-layer networks on the same set of nodes where no notion of inter-layer edges is present in the data.
For changing interactions among the same set of entities over time, we incorporate the description of dynamic centralities \cite{grindrod2011communicability,grindrod2013matrix,grindrod2014dynamical,chen2017dynamic,fenu2017block,arrigo2017sparse,al2021block,arrigo2021dynamic} formulated in \cite{fenu2017block} into our more general multiplex framework.

Choosing the linear algebraic representation of multilayer networks is a non-trivial task \cite{kivela2014multilayer} and for this paper we choose the supra-adjacency matrix as network representation.
This representation has been successfully used to generalize eigenvector centrality to the case of temporal and multiplex networks \cite{taylor2017eigenvector,taylor2019supracentrality,taylor2021tunable}.
However, various other possible network representations exist in the form of different matrix or tensor formulations \cite{kivela2014multilayer,boccaletti2014structure}.
Examples for multilayer centrality measures using different matrix representations include eigenvector centrality \cite{sola2013eigenvector,taylor2017eigenvector,taylor2019supracentrality,taylor2021tunable} and matrix function-based centralities for dynamic networks \cite{grindrod2011communicability,chen2017dynamic,fenu2017block,arrigo2017sparse,benzi2020matrix,al2021block}.
Furthermore, eigenvector centrality for multilayer networks has been defined in terms of third- \cite{tudisco2018node} and fourth-order  \cite{de2015ranking,sole2016random,wang2017identifying,wu2019tensor} tensors.
Additionally, the authors of \cite{sole2016random} use fourth-order tensors to define classical random walk centralities for multilayer networks and the authors of \cite{de2015ranking} define Katz centrality in terms of fourth order tensors, which is an example of a matrix function-based centrality measure.
The development of general tensor function-based centrality measures for tensor representations of multilayer networks is an interesting road for future research.

Our choice of the supra-adjacency matrix as network representation allows us to put well-studied methods from numerical linear algebra to new use as it is currently done in many data-driven applications \cite{stoll2020literature}.
In particular, Krylov subspace methods for the approximation of matrix functions provide highly efficient computational means to evaluate matrix function-based centrality measures even for large-scale problems \cite{saad2003iterative,higham2008functions,golub2013matrix}.
We will discuss that under the assumption of sparsity in the supra-adjacency matrix the runtime of all presented centrality measures scales linearly in the network size.
A particularly elegant technique, which can be applied to compute lower and upper bounds on certain matrix function-based centralities uses the connection between Gauss quadrature, the Lanczos method, and orthogonal polynomials discussed by Golub and Meurant \cite{golub1969calculation,golub1994matrices,golub1997matrices,golub2009matrices}.
Besides network science applications, this technique has been applied to classical numerical linear algebra problems, cf.~e.g., \cite{benzi1999bounds,golub2008approximation}.
For the computation of some matrix function-based centralities, however, the evaluation of a separate matrix function expression for each entity is required, which becomes computationally infeasible for medium to large-scale problems.
To this end, we employ existing numerical techniques for the stochastic and deterministic approximation of the trace and the diagonal of matrix functions \cite{hutchinson1989stochastic,bekas2007estimator,staar2016stochastic,ubaru2017fast,cortinovis2021randomized,meyer2021hutch++}.
Many of the presented computational methods have already been successfully applied to the evaluation of matrix function-based centrality measures on single-layer networks \cite{benzi2020matrix}.

The remainder of this paper is organized as follows.
In Sec.~\ref{sec:Graph_representation} we introduce layer-coupled multiplex networks including their supra-adjacency matrix representation.
Sec.~\ref{sec:Matrix_centralities} introduces existing matrix function-based centrality measures in the context of single-layer networks.
In Sec.~\ref{sec:Multilayer_centralities} we generalize these centrality measures to the case of layer-coupled multiplex networks and employ aggregation techniques, which allow the ranking of nodes, layers, and node-layer pairs.
Sec.~\ref{sec:Numerical_methods} summarizes efficient numerical methods for the approximation of all introduced centrality measures for both directed and undirected as well as weighted and unweighted networks.
Finally, Sec.~\ref{sec:Numerical_experiments} presents extensive numerical experiments on synthetic and real-world networks with the number of layers ranging from $3$ to $124$ and the number of physical nodes ranging from $4$ to $245\,757$.

\section{Multiplex network representation}\label{sec:Graph_representation}

We consider multilayer networks with one aspect (dimension), which consist of a set of $L$ single-layer networks $\Gl = (\Vl, \El),~l=1, \dots , L$ with $\Vl$ denoting the \emph{vertex sets} and $\El \subset \Vl \times \Vl$ the \emph{edge sets} of layers $l=1, \dots , L$ \cite{kivela2014multilayer}.
The different \emph{layers} represent different kinds of interactions or relationships between its entities or changing interactions over time.
We call $x_i$ a \emph{physical node}, which is represented by different instances $\xl_i \in \Vl$ of itself in the different layers called \emph{node-layer pairs}.
We assume all layers to be node-aligned, i.e., to consist of the common vertex set $\tilde{\V} = \V^{(1)} = \dots = \V^{(L)}$.
This situation can always be enforced by adding isolated nodes where necessary.

We distinguish \emph{intra-layer edges} connecting node-layer pairs from the same layer and \emph{inter-layer edges} connecting node-layer-pairs from different layers.
In this paper, all edges can be positively weighted or unweighted as well as directed or undirected.

We define intra-layer edge weights by weight functions $\wl: \Vl \times \Vl \rightarrow \R_{\geq 0}$, not necessarily equal for all $l=1 , \dots , L$.
In the unweighted case, we define 
\begin{equation*}
	\wl_{ij}=
	\begin{cases}
		1 & \text{if }(\xl_i, \xl_j) \in \El, \\
		0 &  \text{otherwise.}
	\end{cases}
\end{equation*}
In the weighted case, we extend the definition to allow $\wl_{ij} \in \R_{>0}$ if $(\xl_i, \xl_j) \in \El$.
In either case, we define the \emph{single-layer adjacency matrices} $\Al \in \R^{n \times n}_{\geq 0}$ for $l=1 , \dots , L$ via
\begin{equation}\label{eq:single_layer_adjacencies}
	\Al_{ij} = \wl_{ij}.
\end{equation}
Note that we have $(\Al)^T=\Al$ for undirected layers and $(\Al)^T\neq\Al$ when layer $l$ contains at least one directed intra-layer edge.

\begin{figure}
	\centering
	\includegraphics[width=0.4\textwidth]{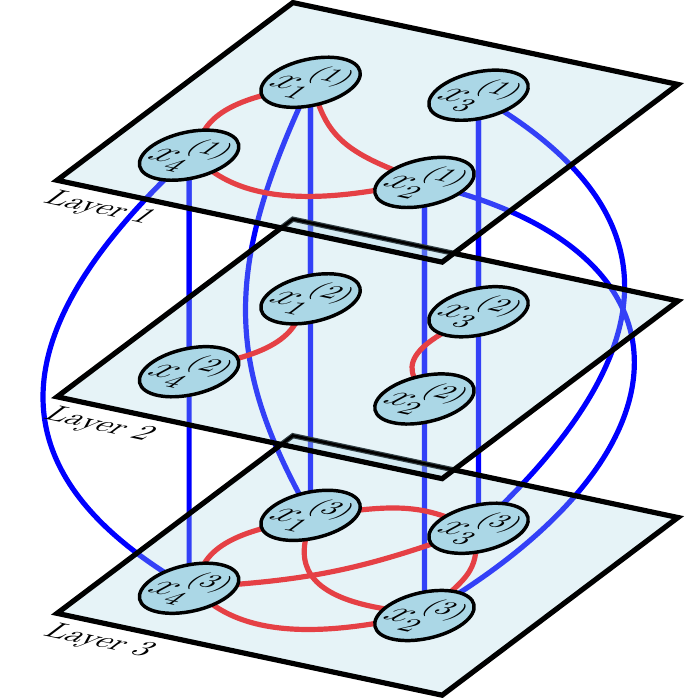}
	\caption{Example of a layer-coupled multiplex network with $4$ nodes and $3$ layers.
	Intra-layer edges are marked red and inter-layer edges are marked blue.}\label{fig:network_examples_multi}
\end{figure}

Additionally, we define an inter-layer edge set $\tilde{\E} \subset \Vl \times \Vk$ containing inter-layer edges $(\xl_i, \xk_j)$ where $l \neq k$.
We restrict ourselves to layer-coupled multiplex networks in which each node-layer pair is only allowed to form an inter-layer edge to instances of its physical node in other layers, i.e., we only allow inter-layer edges $(\xl_i, \xk_j)$ with $l \neq k$ and $i = j$.
Furthermore, we fix an inter-layer edge weight $\tilde{\bm{A}}_{lk}\geq 0$ for each pair of layers $l$ and $k$ and we collect these weights in the \emph{inter-layer weight matrix} $\tilde{\bm{A}}\in\R^{L \times L}_{\geq 0}$.
The layer coupling can then be represented by the Kronecker product $\tilde{\bm{A}} \otimes \bm{I}$ with the identity matrix $\bm{I}\in\R^{n \times n}$.
We have $\tilde{\bm{A}}^T = \tilde{\bm{A}}$ in the case of undirected inter-layer edges and $\tilde{\bm{A}}^T \neq \tilde{\bm{A}}$ if at least one inter-layer edge is directed.
The applicability of the methods introduced in this paper to more general inter-layer edges for the case of undirected networks is studied in \cite{bergermann2021orientations}.

With the above definitions, we define (node-aligned) \emph{layer-coupled multiplex networks} $\G=(\tilde{\V}, \E^{(1)}, \dots , \E^{(L)}, \tilde{\E})$ consisting of the common vertex set $\tilde{\V}$, intra-layer edge sets $\E^{(1)}, \dots , \E^{(L)}$, and the inter-layer edge set $\tilde{\E}$.
There is a multitude of different multilayer network representations available in the literature including matrix as well as third- and fourth-order tensor representations, cf.\ \cite{de2015ranking,sole2016random,wang2017identifying,tudisco2018node,wu2019tensor} as well as \cite{kivela2014multilayer} for an overview.
For this paper, we choose a supra-adjacency matrix representation \cite[Sec.~2.3]{kivela2014multilayer}.
Note that this representation corresponds to a special case of the supracentrality matrix that has been used in \cite{taylor2017eigenvector,taylor2019supracentrality,taylor2021tunable} to generalize eigenvector centrality to multiplex networks.

The \emph{supra-adjacency matrix} $\bm{A}\in\R^{nL \times nL}_{\geq 0}$ is defined as the weighted sum of a \emph{multilayer intra-layer adjacency matrix} $\Aintra\in\R^{nL \times nL}_{\geq 0}$ containing the individual layer adjacency matrices on its block diagonal and an \emph{inter-layer adjacency matrix} $\Ainter\in\R^{nL \times nL}_{\geq 0}$ representing the layer-coupling, i.e.,
\begin{align}
	\bm{A} & = \Aintra + \omega \Ainter \notag \\ 
	& = \text{blkdiag}[\A1, \dots , \AL] + \omega \bm{\tilde{A}} \otimes \bm{I} \notag \\
	& = 
	\begin{bmatrix}
		\A1 & \dots & \bm{0}\\
		\vdots & \ddots & \vdots\\
		\bm{0} & \dots & \AL
	\end{bmatrix}
	+ \omega 
	\begin{bmatrix}
		\bm{\tilde{A}}_{11} \bm{I} & \dots & \bm{\tilde{A}}_{1L} \bm{I}\\
		\vdots & \ddots & \vdots\\
		\bm{\tilde{A}}_{L1} \bm{I} & \dots & \bm{\tilde{A}}_{LL} \bm{I}\\
	\end{bmatrix},\label{eq:supra_adjacency}
\end{align}
where $\bm{0}\in\R^{n \times n}$ denotes the zero matrix and the \emph{coupling parameter} $\omega\geq 0$ controls the relative importance of the two types of edges.
We discuss possible choices for the inter-layer weight matrix $\tilde{\bm{A}}$ in Sec.~\ref{sec:Multilayer_centralities}.
Fig.~\ref{fig:network_examples_multi} shows an example layer-coupled multiplex network with $n=4$ physical nodes and $L=3$ layers.
Note that we have $\bm{A}^T=\bm{A}$ if all $\Al$ and $\tilde{\bm{A}}$ are symmetric and $\bm{A}^T \neq \bm{A}$ if at least one (intra- or inter-layer) edge is directed.
Note also that from a computational point of view $\bm{A}$ does not have to be formed explicitly as all information is encoded in $\Al, l=1, \dots , L$ as well as $\tilde{\bm{A}}$.

\section{Matrix function-based centrality measures}\label{sec:Matrix_centralities}

There is a multitude of centrality measures available in the literature, which identify and rank the most central nodes of a complex network, cf.~e.g., \cite{freeman1977set,freeman1978centrality,bonacich1987power,brin1998anatomy,page1999pagerank,kleinberg1999authoritative,sola2013eigenvector,gomez2013diffusion,de2014navigability,de2015ranking,gleich2015pagerank,sole2016random,taylor2017eigenvector,wang2017identifying,tudisco2018node,taylor2019supracentrality,benson2019three,wu2019tensor,taylor2021tunable}.
In this section, we motivate and define matrix function-based centrality measures, which have been intensively studied for single-layer \cite{katz1953new,estrada2000characterization,estrada2005subgraph,estrada2010network,benzi2013ranking,benzi2020matrix} and dynamic networks \cite{grindrod2011communicability,grindrod2013matrix,grindrod2014dynamical,chen2017dynamic,fenu2017block,arrigo2017sparse,al2021block,arrigo2021dynamic} in recent years at the example of undirected and unweighted single-layer networks.
In this situation, we have $\bm{A}=\A1\in\R^{n \times n}_{\geq 0}$ and we drop the superscript $(l)$ indicating the layer ID for all related quantities in the notation throughout this section.

\begin{figure}
	\centering
	\subfloat[]{
		{\setlength{\fboxsep}{0pt}\fbox{\includegraphics[width=.14\textwidth]{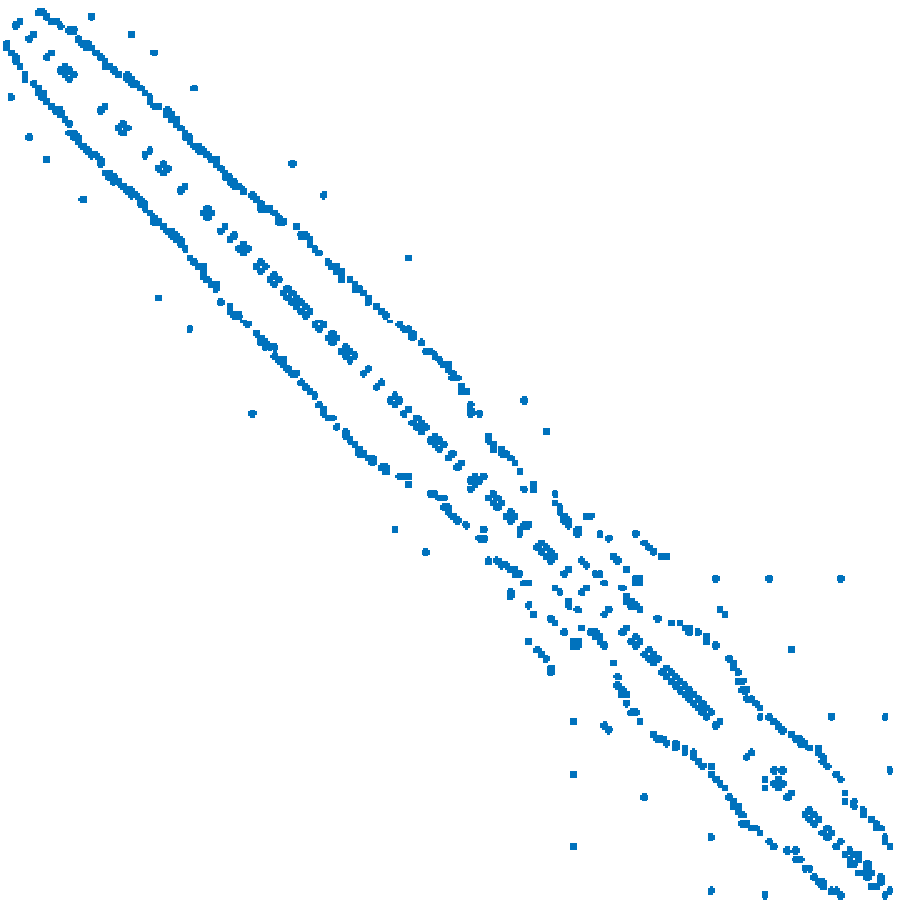}}}
	}
	\hfill
	\subfloat[]{
		{\setlength{\fboxsep}{0pt}\fbox{\includegraphics[width=.14\textwidth]{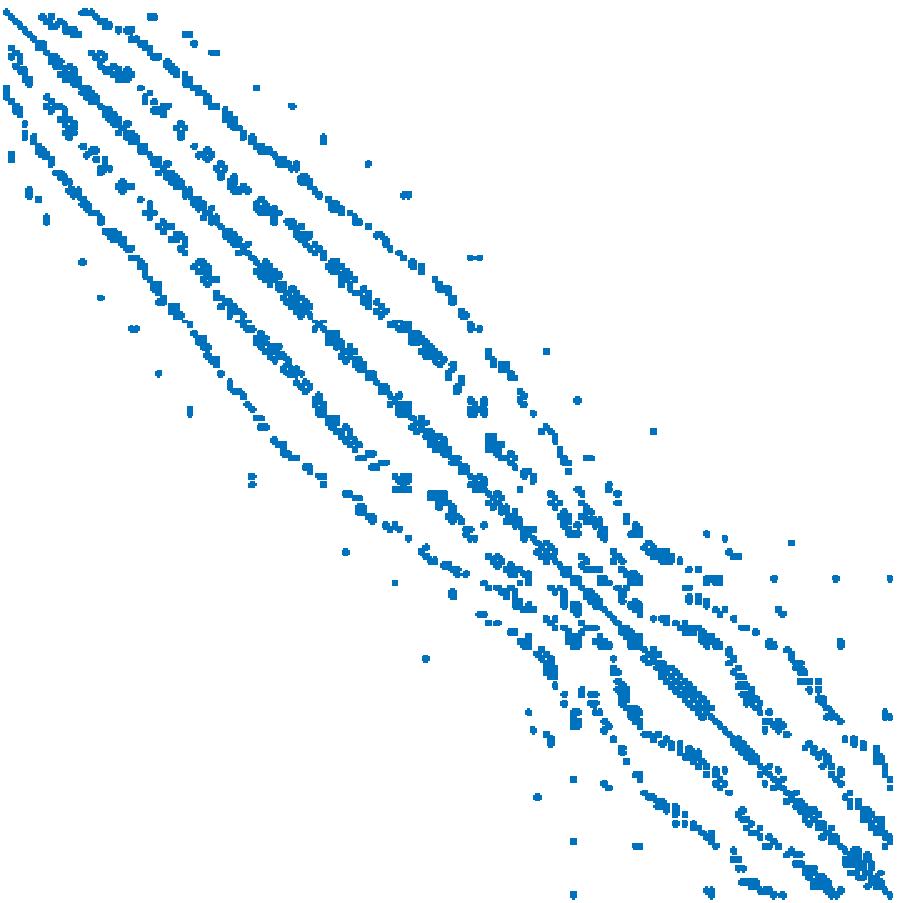}}}
	}
	\hfill
	\subfloat[]{
		{\setlength{\fboxsep}{0pt}\fbox{\includegraphics[width=.14\textwidth]{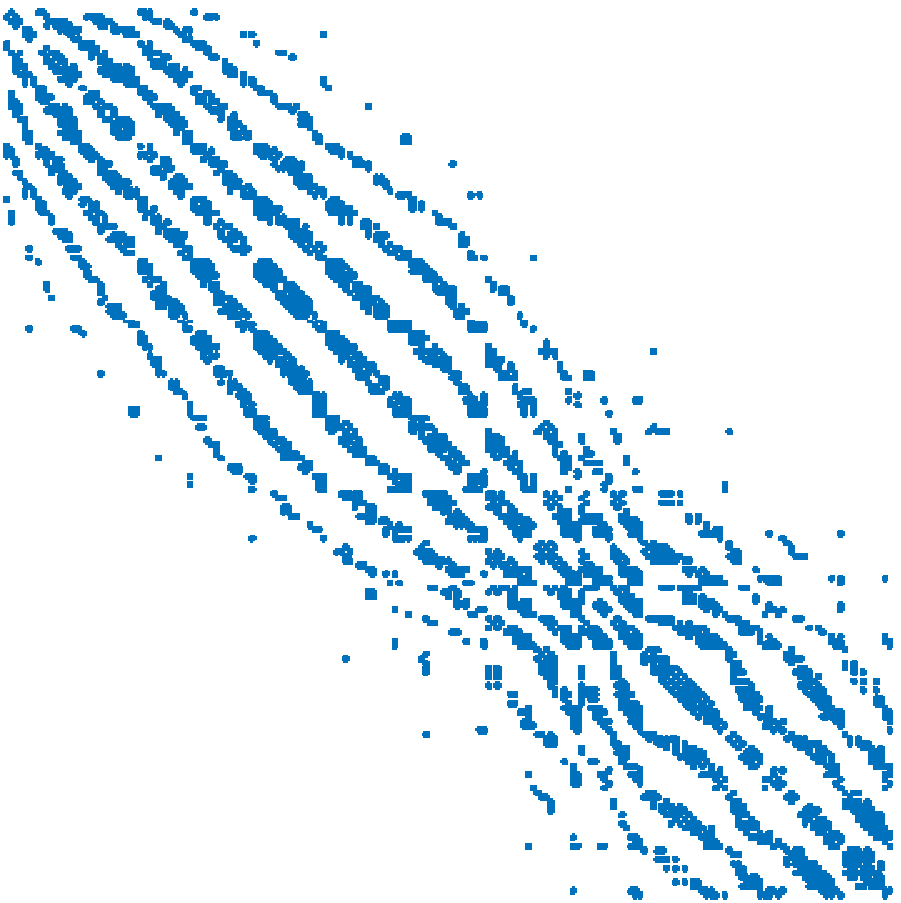}}}
	}
	\\
	\subfloat[]{
		{\setlength{\fboxsep}{0pt}\fbox{\includegraphics[width=.14\textwidth]{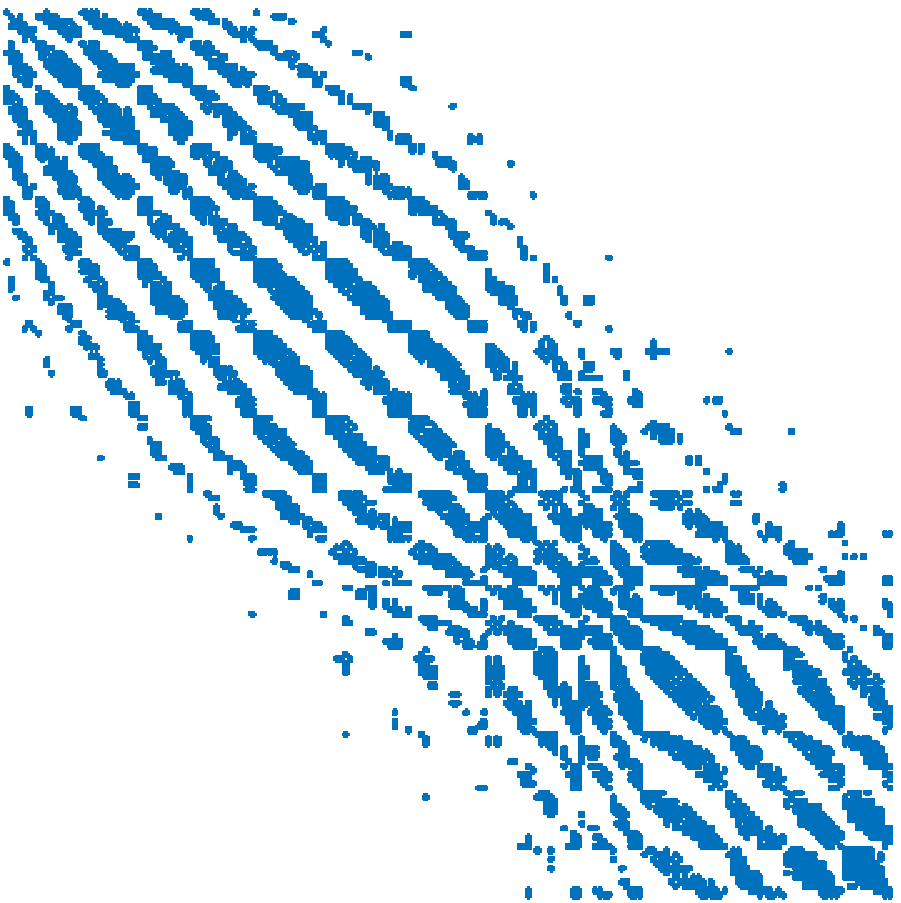}}}
	}
	\hfill
	\subfloat[]{
		{\setlength{\fboxsep}{0pt}\fbox{\includegraphics[width=.14\textwidth]{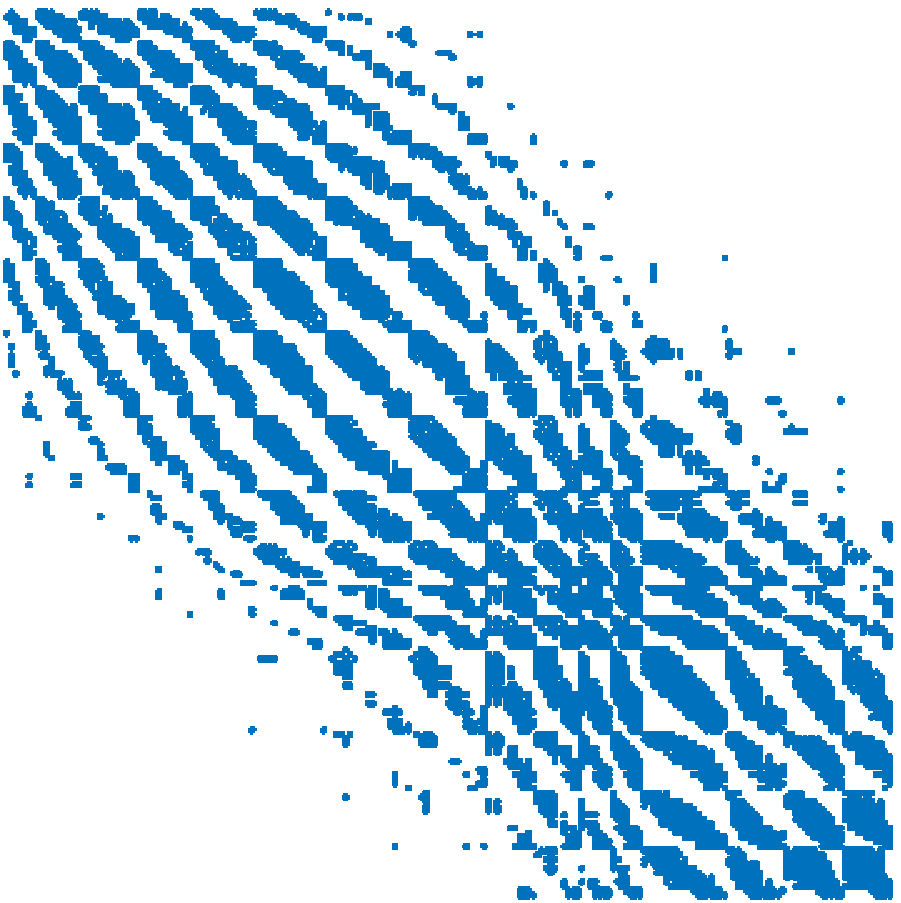}}}
	}
	\hfill
	\subfloat[]{
		{\setlength{\fboxsep}{0pt}\fbox{\includegraphics[width=.14\textwidth]{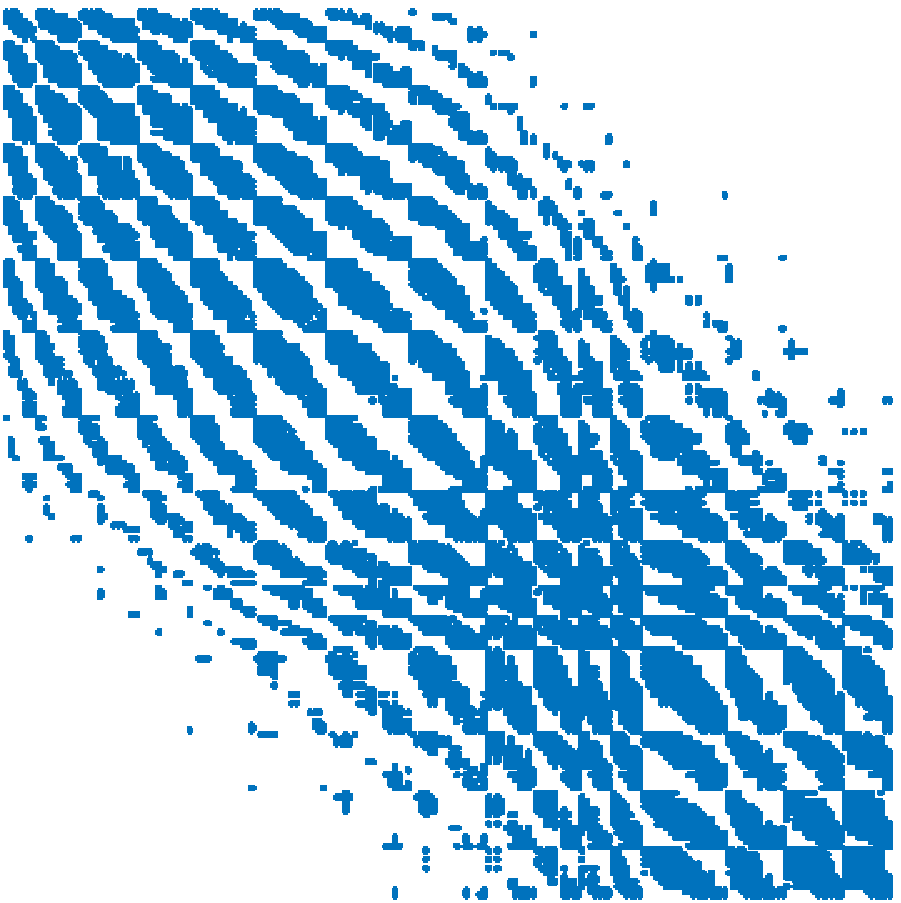}}}
	}
	\\
	\subfloat[]{
		{\setlength{\fboxsep}{0pt}\fbox{\includegraphics[width=.14\textwidth]{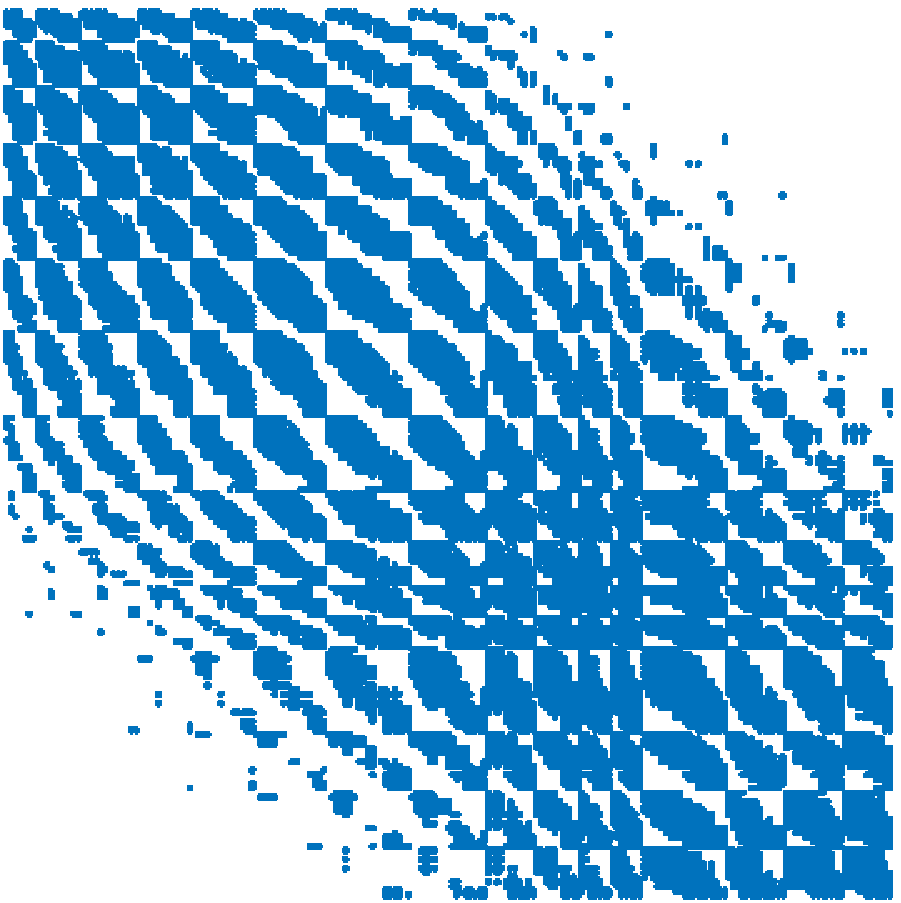}}}
	}
	\hfill
	\subfloat[]{
		{\setlength{\fboxsep}{0pt}\fbox{\includegraphics[width=.14\textwidth]{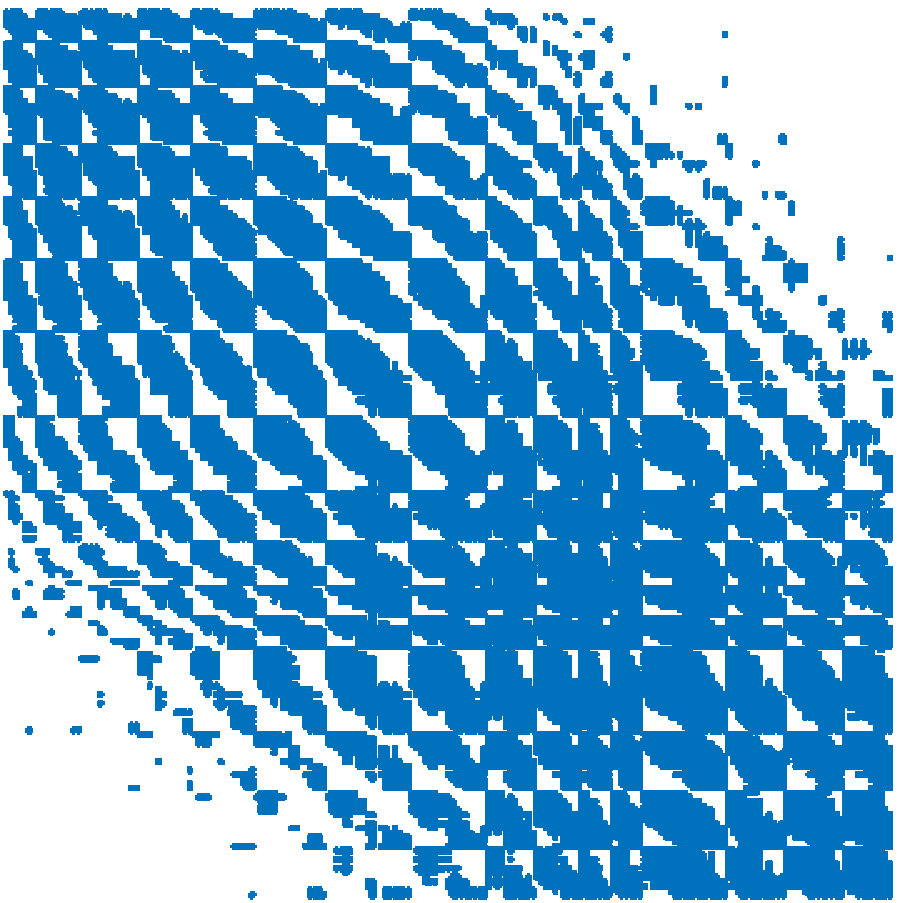}}}
	}
	\hfill
	\subfloat[]{
		{\setlength{\fboxsep}{0pt}\fbox{\includegraphics[width=.14\textwidth]{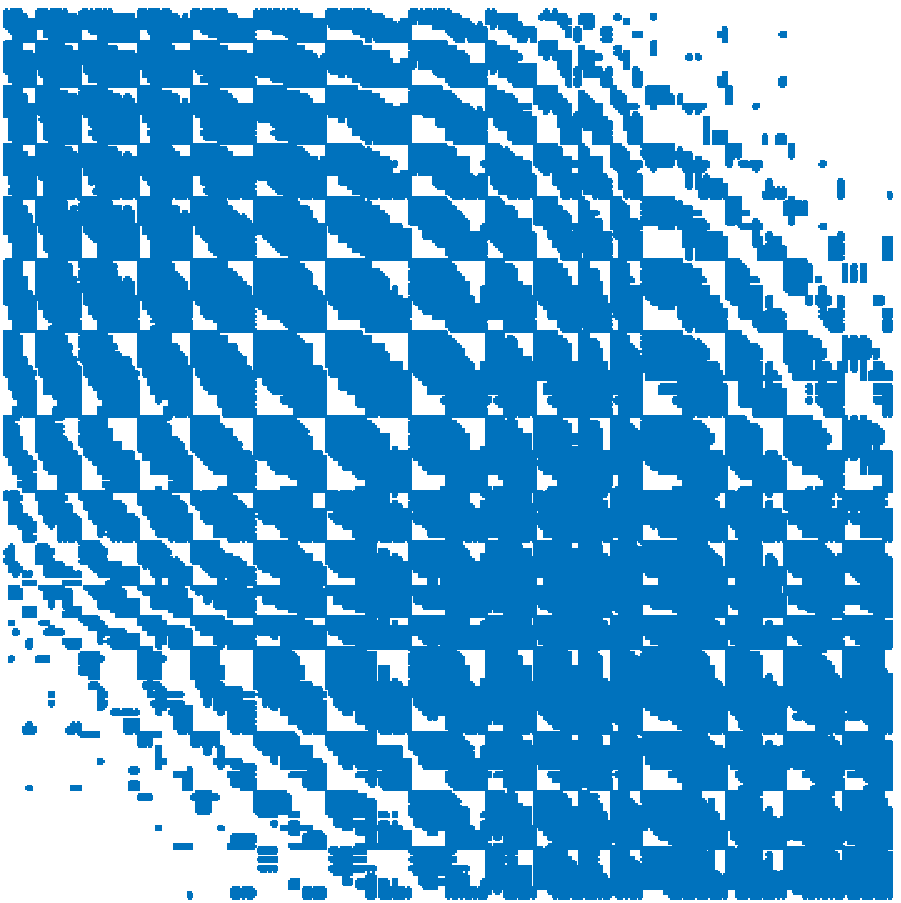}}}
	}
	\caption{Sparsity structure of the first nine adjacency matrix powers $\bm{A}, \dots , \bm{A}^{9}$ for the taxi layer of the Scotland Yard network, which is a connected undirected single-layer network with $n=199$ nodes, cf.~Sec.~\ref{sec:Numerical_experiments_approximation_error}.
	The quantity $\rho$ denotes the matrix density, i.e., the proportion of non-zero entries in the respective matrix power $\bm{A}^p\in\R^{199 \times 199}_{\geq 0}$.
	(a) $\bm{A}, \rho=0.017$, (b) $\bm{A}^2, \rho=0.044$, (c) $\bm{A}^3, \rho=0.091$, (d) $\bm{A}^4, \rho=0.16$, (e) $\bm{A}^5, \rho=0.245$, (f) $\bm{A}^6, \rho=0.338$, (g) $\bm{A}^7, \rho=0.434$, (h) $\bm{A}^8, \rho=0.531$, (i) $\bm{A}^9, \rho=0.626$.
	}\label{im:sy_powers_single}
\end{figure}

It is well-known from graph theory that for an unweighted and undirected network the entry $[\bm{A}^k]_{ij}$ denotes the number of walks of length $k$ existing between nodes $x_i$ and $x_j$ \cite{estrada2012structure}.
A walk of length $k$ is defined by a sequence of $k$ adjacent nodes, which may contain repeated nodes, i.e., we allow backtracking walks.
For non-backtracking walks we refer the reader to \cite{alon2007non,arrigo2018exponential,arrigo2018non}.
In the special case $i=j$, we speak of closed walks that start and end at node $x_i$.
These entries correspond to the diagonal elements of the adjacency matrix powers.
Fig.~\ref{im:sy_powers_single} illustrates the sparsity structure of the adjacency matrix powers $\bm{A}, \dots , \bm{A}^{9}$ for the taxi layer of the Scotland Yard network, which is a connected undirected single-layer network with $n=199$ nodes, cf.~Sec.~\ref{sec:Numerical_experiments_approximation_error} for details.

Adjacency matrix powers can, e.g., be used to define \emph{degree} and \emph{eigenvector centrality}.
The degree centrality of node $x_i$ is given by $d_i=\bm{e}_i^T \bm{A} \bm{1}$ where $\bm{e}_i \in \R^n$ denotes the $i$th unit vector and $\bm{1} = [1 , \dots , 1]^T \in \R^n$ the one vector.
Eigenvector centrality is defined by the entries of the eigenvector $\bm{\phi}$ corresponding to the largest eigenvalue $\lmax$ of $\bm{A}$.
Under mild conditions, this can, e.g., be obtained by a power iteration approximating the limit $\lim_{p\rightarrow \infty} \bm{A}^p \bm{v}/\| \bm{A}^p \bm{v} \|_2$ with a suitable starting vector $\bm{v}\in\R^n$.
Degree centrality can be viewed as a local measure taking only direct neighbors into account while eigenvector centrality depicts a global measure representing the stationary distribution of walkers on the network.

The idea of matrix function-based centrality measures is to interpolate between local degree and global eigenvector centrality by considering walks of all lengths (or subgraphs of all sizes).
This idea is formalized by the adjacency matrix power series $\sum_{p=0}^{\infty} \bm{A}^p$.
A number of works by Estrada and co-authors \cite{estrada2000characterization,estrada2005subgraph,estrada2008communicability,estrada2010network} developed different scaling mechanisms, which assign less weight to longer walks leading to the power series of the frequently used matrix exponential function
\begin{equation}\label{eq:matrix_exponential}
	\sum_{p=0}^{\infty} \frac{\beta^p}{p!} \bm{A}^p = \bm{I} + \beta \bm{A} + \frac{\beta^2}{2} \bm{A}^2 + \frac{\beta^3}{3!} \bm{A}^3 + \dots = e^{\beta \bm{A}},
\end{equation}
with the inverse temperature $\beta>0$ \cite{estrada2008communicability} as well as the matrix resolvent function
\begin{equation}\label{eq:matrix_resolvent}
	\sum_{p=0}^{\infty} \alpha^p \bm{A}^p= \bm{I} + \alpha \bm{A} + \alpha^2 \bm{A}^2 + \dots = (\bm{I}-\alpha \bm{A})^{-1},
\end{equation}
which is convergent for $0<\alpha<1/\lmax$ \cite{estrada2010network}.

The centrality measures considered in this paper \cite{benzi2013total,katz1953new,estrada2005subgraph,estrada2010network,estrada2000characterization,estrada2008communicability} can be derived from certain matrix function expressions, which are summarized in Tab.~\ref{tab:categories_measures}.
The diagonal entries of the matrix functions $\bm{e}_i^T f(\bm{A}) \bm{e}_i$ can be viewed as the weighted sum of closed walks starting and ending at node $x_i$.
Similarly, the \emph{communicability} between two nodes $x_i$ and $x_j$ is measured by the weighted sum of walks starting at $x_i$ and ending at $x_j$.
Entries of the row sum vector of the matrix functions $\bm{e}_i^T f(\bm{A}) \bm{1}$, in turn, count all walks starting at node $x_i$ regardless of the end point of the walk.
Finally, the \emph{Estrada index} and \emph{total network communicability} provide scalar measures for the connectivity of the full network.

\begin{table}
	\begin{center}
		\begin{tabular}{c @{\hspace{20pt}} c}
			\hline \hline
			\multirow{4}{25pt}{$f(\bm{A}) \bm{b}$} & Total communicability \cite{benzi2013total}\\
			 & $TC(i,\beta)=\bm{e}_i^T e^{\beta \bm{A}} \bm{1}$\\
			 & Katz centrality \cite{katz1953new}\\
			 & $KC(i,\alpha)=\bm{e}_i^T (\bm{I}-\alpha \bm{A})^{-1} \bm{1}$\\ \hline
			\multirow{8}{38pt}{$\bm{u}^T f(\bm{A}) \bm{u}$} & Subgraph centrality \cite{estrada2005subgraph}\\
			 & $SC(i,\beta)=\bm{e}_i^T e^{\beta \bm{A}} \bm{e}_i$\\
			 & Resolvent-based subgraph centrality \cite{estrada2010network}\\
			 & $SC_{\mathrm{res}}(i,\alpha)=\bm{e}_i^T (\bm{I}-\alpha \bm{A})^{-1} \bm{e}_i$\\
			 & Estrada index \cite{estrada2000characterization}\\
			 & $EI(\G,\beta)=\sum_{i=1}^{nL} \bm{e}_i^T e^{\beta \bm{A}} \bm{e}_i$\\
			 & Total network communicability \cite{benzi2013total}\\
			 & $TNC(\G,\beta)=\frac{1}{nL} \bm{1}^T e^{\beta \bm{A}} \bm{1}$\\ \hline
			\multirow{2}{38pt}{$\bm{u}^T f(\bm{A}) \bm{v}$} & Communicability \cite{estrada2008communicability}\\
			 & $C(i,j,\beta)=\bm{e}_i^T e^{\beta \bm{A}} \bm{e}_j$\\ \hline \hline
		\end{tabular}
	\end{center}
	\caption{Overview of all defined matrix function-based centrality measures categorized into more general matrix function expressions.
	Note that throughout Sec.~\ref{sec:Matrix_centralities} we have assumed $L=1$.}\label{tab:categories_measures}
\end{table}

In the case of weighted adjacency matrices, the entries in $[\bm{A}^k]_{ij}$ can no longer be interpreted as the number of walks of length $k$ between nodes $x_i$ and $x_j$.
However, formally all above definitions equally apply and the elements of the matrix powers still contain information about the relative connectivity of pairs of nodes.

In the case of directed networks, i.e., $\bm{A}^T\neq\bm{A}$ we must distinguish between each node's role as broadcaster and receiver.
Similarly to degree and eigenvector centrality we obtain \emph{broadcaster centralities} with the definitions from Tab.~\ref{tab:categories_measures} and \emph{receiver centralities} by replacing $\bm{A}$ by $\bm{A}^T$ \cite{benzi2013ranking}.
However, subgraph centrality and resolvent-based subgraph centrality, which are defined as the diagonal elements $[f(\bm{A})]_{ii}$, can not differentiate between broadcaster and receiver centrality as by \cite[Thm.~1.13(b)]{higham2008functions} we have $f(\bm{A}^T)=f(\bm{A})^T$ and thus $[f(\bm{A})]_{ii}=[f(\bm{A}^T)]_{ii}$ for all $i=1, \dots ,n$.
In this case, we can instead consider the \emph{symmetric bipartite representation} of a directed network \cite{benzi2013ranking}, which is defined as
\begin{equation}\label{eq:bipartite_adjacency}
	\bm{\mathcal{A}}=
	\begin{bmatrix}
	\bm{0} & \bm{A}\\\bm{A}^T & \bm{0}
	\end{bmatrix}
	\in \R^{2n \times 2n}_{\geq 0},
\end{equation}
and obtain broadcaster centralities as $[f(\bm{\mathcal{A}})]_{ii}$ for $i=1, \dots ,n$ and receiver centralities as $[f(\bm{\mathcal{A}})]_{ii}$ for $i=n+1, \dots ,2n$ for both $f(\bm{\mathcal{A}})=e^{\beta \bm{\mathcal{A}}}$ and $f(\bm{\mathcal{A}})=(\bm{I} - \alpha \bm{\mathcal{A}})^{-1}$.

\section{Definition of multiplex matrix function-based centrality measures}\label{sec:Multilayer_centralities}

We now generalize the matrix function-based centrality measures introduced for single-layer networks in Sec.~\ref{sec:Matrix_centralities} to the case of layer-coupled multiplex networks.
To this end, we return to the multiplex network representation specified in Sec.~\ref{sec:Graph_representation} where each vertex $\xl_i$ represents a node-layer pair, i.e., the instance of physical node $x_i$ in layer $l$.
Consequently, we add layer indices to all quantities in Tab.~\ref{tab:categories_measures}, e.g., $KC(i,l,\alpha)$ or $C(i,l,j,k,\beta)$.

We propose to extend the interpretation of adjacency matrix powers from the single-layer case to the supra-adjacency matrix defined in Eq.~\eqref{eq:supra_adjacency}.
As this contains information about both intra- and inter-layer edges one step of a walk on the multiplex network starting from node $\xl_i$ can either follow an intra-layer edge towards $\xl_j$ within the same layer or an inter-layer edge towards $\xk_i$ in another layer, given that at least one such edge exists for $i \neq j$ or $l \neq k$, respectively.
Consequently, both intra- and inter-layer edge weights need to reflect the connectivity between node-layer pairs in the multiplex network model.
Depending on the application these weights could, e.g., describe the ability of the network to spread travelers, information, ideas, etc.
The trade-off between intra- and inter-layer edge weights is controlled by the coupling parameter $\omega$.
More examples of walk-based centrality measures that consider walks along intra- and inter-layer edges can, e.g., be found in \cite{gomez2013diffusion,de2014navigability,sole2016random}.

Available knowledge from the particular application about the coupling strength of the layers encoded in the inter-layer weight matrix $\tilde{\bm{A}}$ increases the quality of the network model.
However, if no such information is present in the data some standard inter-layer couplings can be used to create multiplex networks from several single-layer networks.
Two options of such inter-layer couplings are unweighted all-to-all coupling with and without self-edges, which are represented by $\tilde{\bm{A}} = \bm{1}\bm{1}^T$ and $\tilde{\bm{A}} = \bm{1}\bm{1}^T - \bm{I}$, respectively.
We illustrate in Sec.~\ref{sec:Numerical_experiments_small_synthetic_network} that this choice is better capable to reflect the underlying structure of inherently multilayered networks than aggregated networks.
For temporal networks, we use the block matrix formulation of dynamic centralities \cite{fenu2017block}, which corresponds to $\tilde{\bm{A}}$ being zero except for the super-diagonal.
The latter consists of the weights $\tilde{\bm{A}}_{(l-1),l} = e^{-\Delta t_l}, l=2, \dots , L$ with $\Delta t_l$ the time difference between layers $l-1$ and $l$, which reflects the increased importance of more recent walks.

\begin{figure}
	\centering
	\subfloat[]{
		{\setlength{\fboxsep}{0pt}\fbox{\includegraphics[width=.14\textwidth]{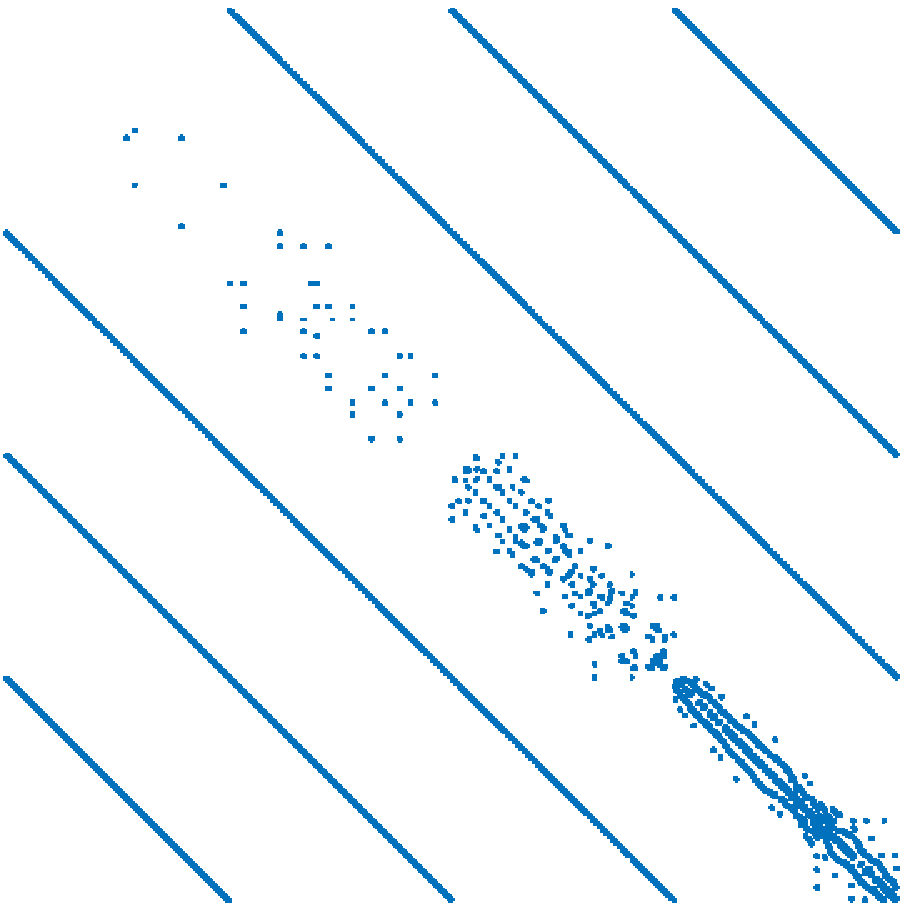}}}
	}
	\hfill
	\subfloat[]{
		{\setlength{\fboxsep}{0pt}\fbox{\includegraphics[width=.14\textwidth]{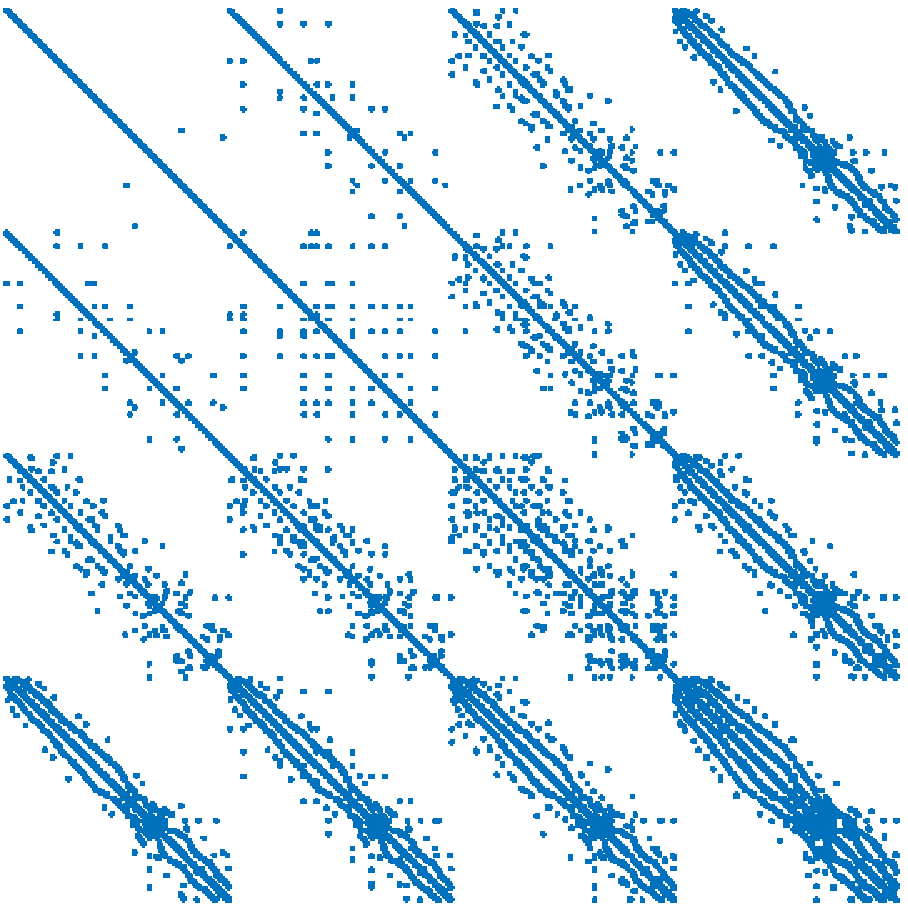}}}
	}
	\hfill
	\subfloat[]{
		{\setlength{\fboxsep}{0pt}\fbox{\includegraphics[width=.14\textwidth]{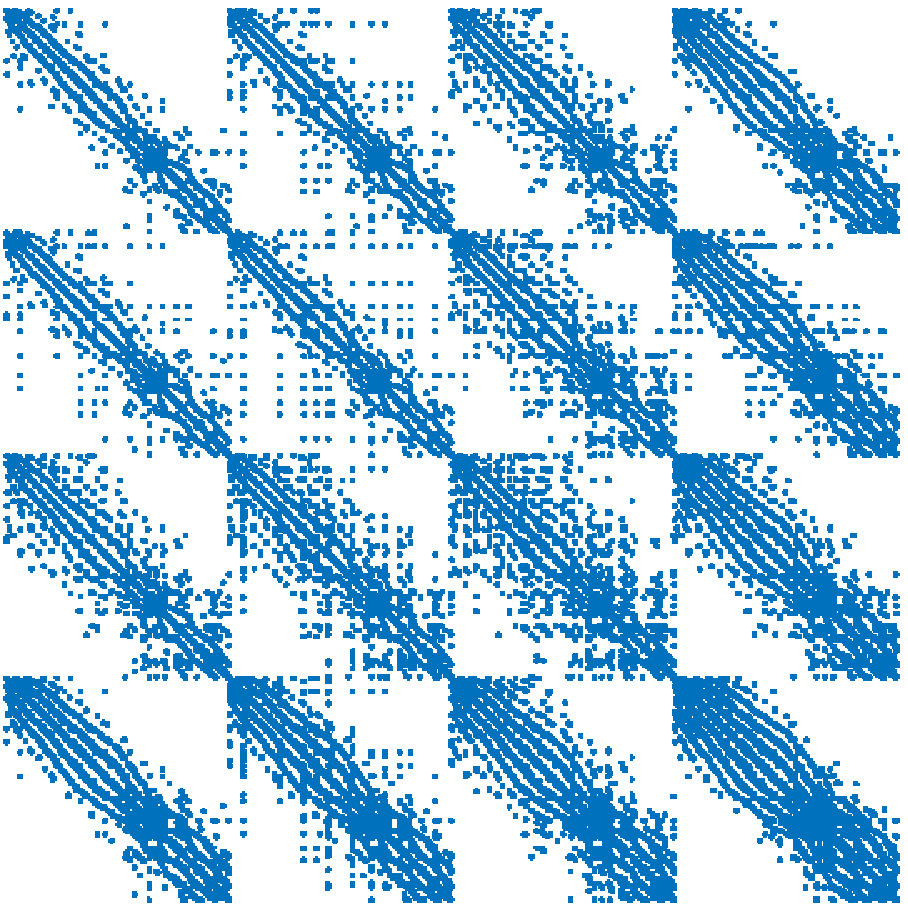}}}
	}
	\\
	\subfloat[]{
		{\setlength{\fboxsep}{0pt}\fbox{\includegraphics[width=.14\textwidth]{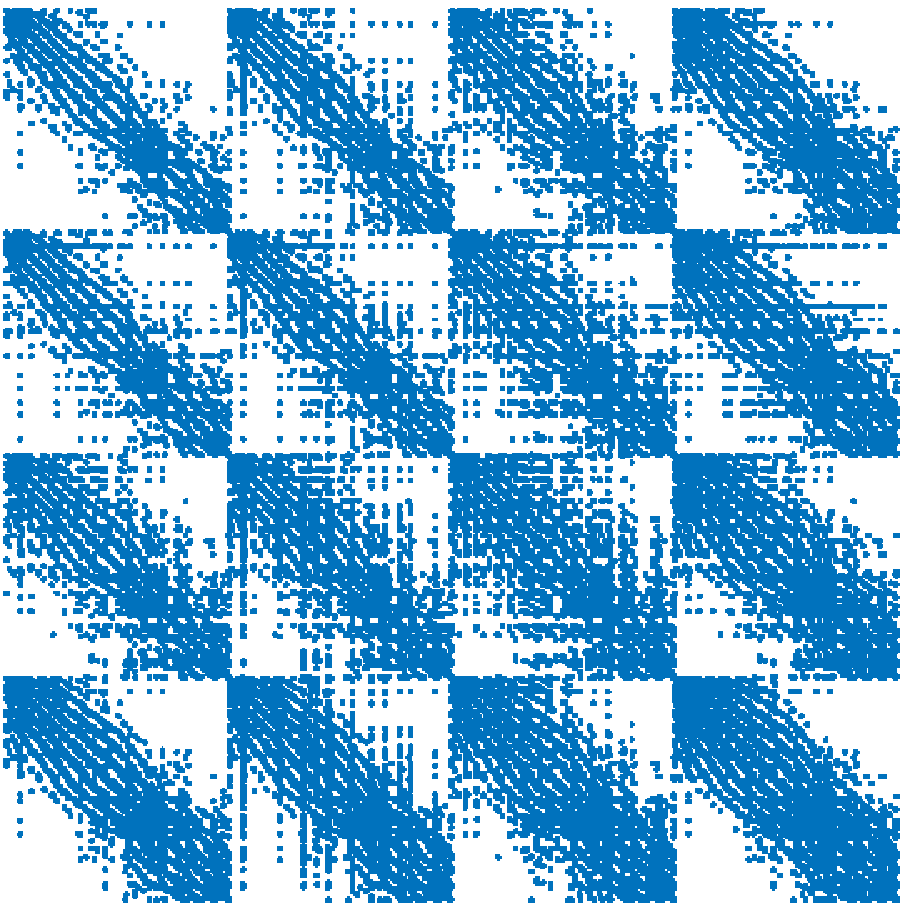}}}
	}
	\hfill
	\subfloat[]{
		{\setlength{\fboxsep}{0pt}\fbox{\includegraphics[width=.14\textwidth]{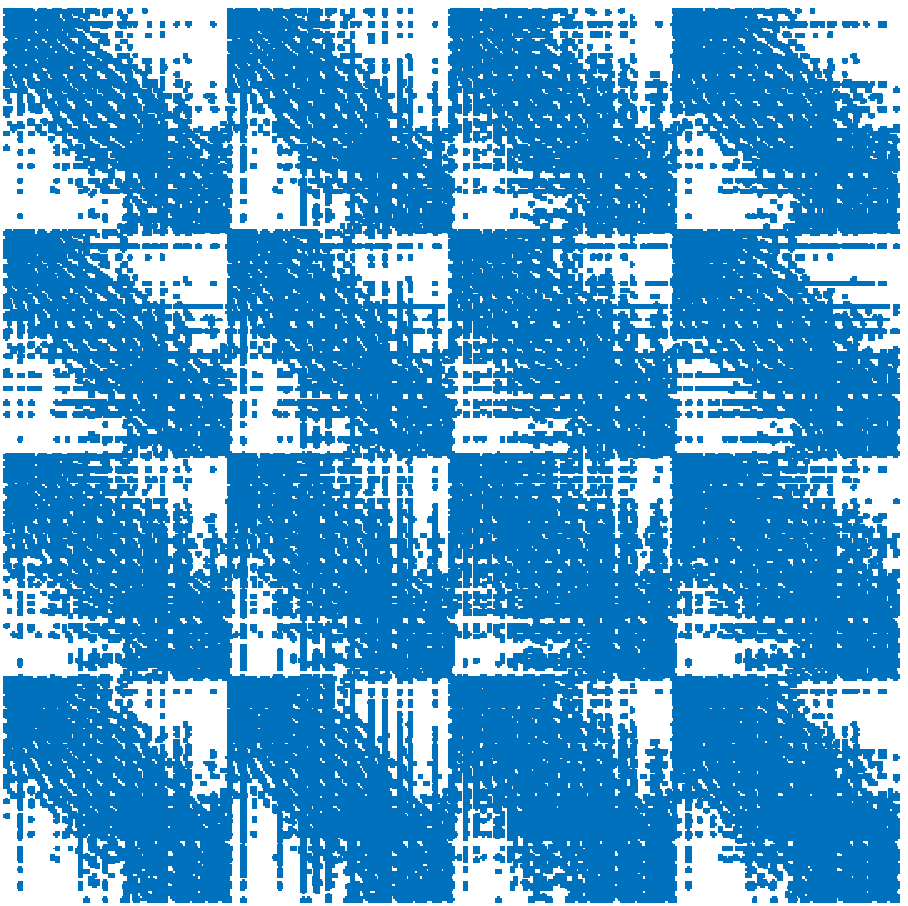}}}
	}
	\hfill
	\subfloat[]{
		{\setlength{\fboxsep}{0pt}\fbox{\includegraphics[width=.14\textwidth]{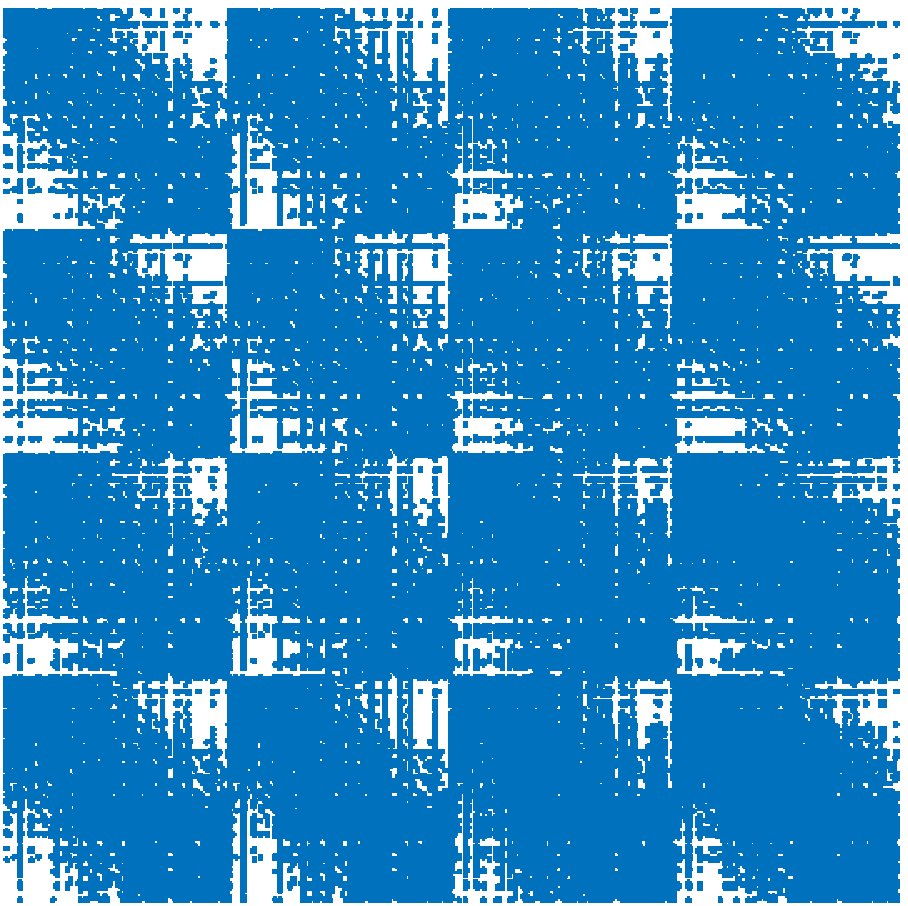}}}
	}
	\\
	\subfloat[]{
		{\setlength{\fboxsep}{0pt}\fbox{\includegraphics[width=.14\textwidth]{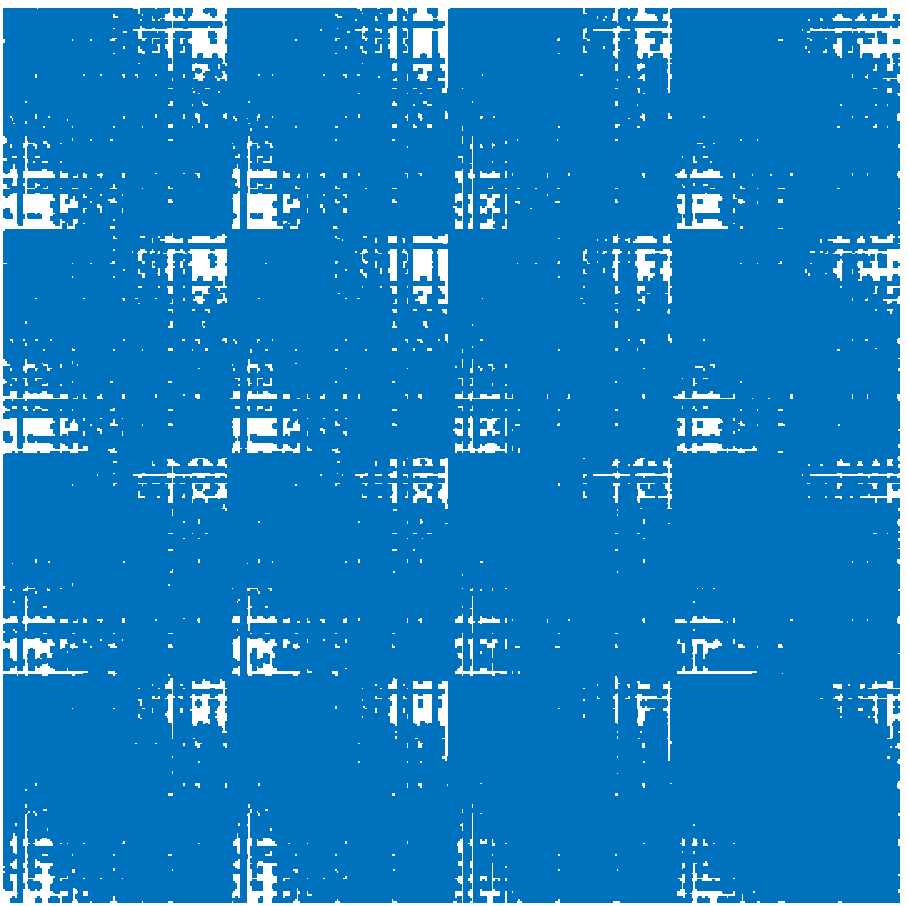}}}
	}
	\hfill
	\subfloat[]{
		{\setlength{\fboxsep}{0pt}\fbox{\includegraphics[width=.14\textwidth]{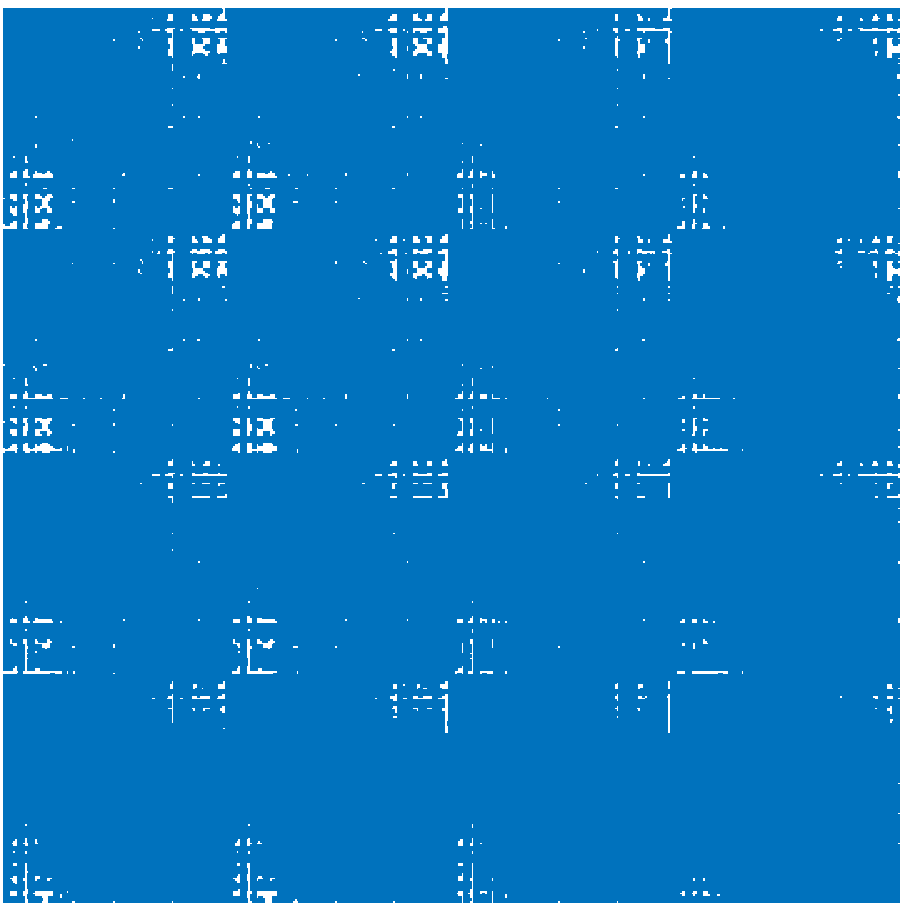}}}
	}
	\hfill
	\subfloat[]{
		{\setlength{\fboxsep}{0pt}\fbox{\includegraphics[width=.14\textwidth]{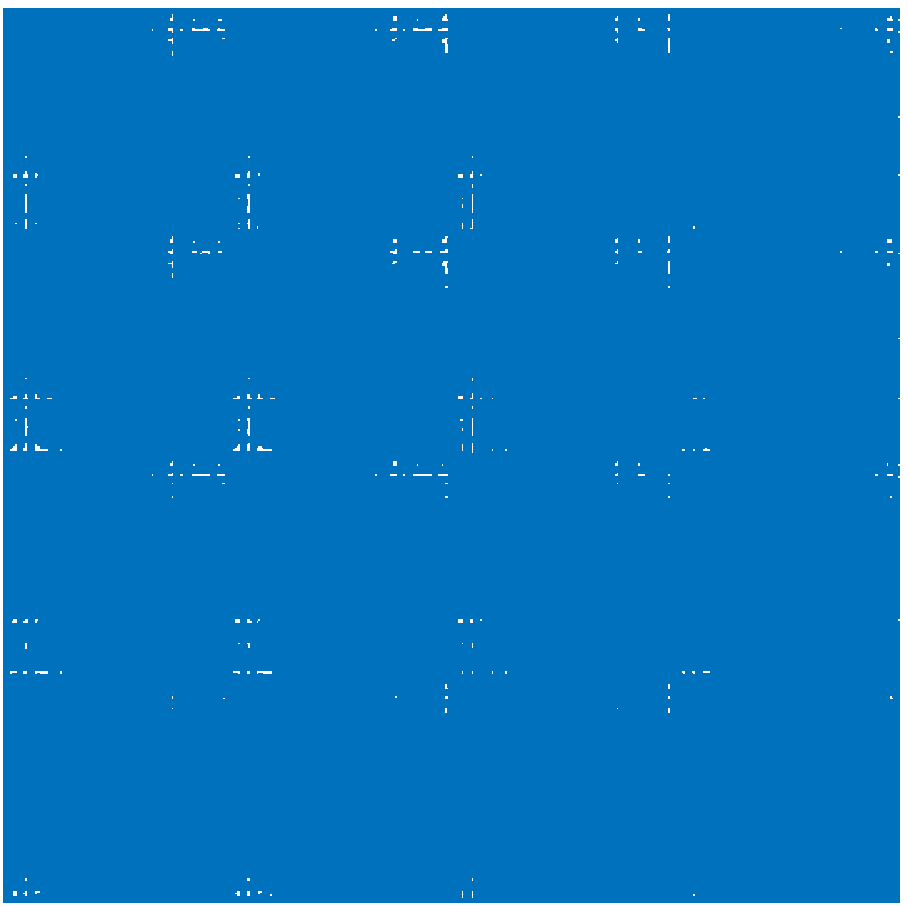}}}
	}
	\caption{Sparsity structure of the first nine supra-adjacency matrix powers $\bm{A}, \dots , \bm{A}^{9}$ of the undirected layer-coupled Scotland Yard multiplex network with $n=199$ nodes and $L=4$ layers, cf.~Sec.~\ref{sec:Numerical_experiments_approximation_error}, as well as all-to-all inter-layer coupling without self-edges, i.e., $\tilde{\bm{A}} = \bm{1}\bm{1}^T - \bm{I}$.
	Layer four in the bottom right block corresponds to the single-layer adjacency matrix from Fig.~\ref{im:sy_powers_single}.
	The quantity $\rho$ denotes the matrix density, i.e., the proportion of non-zero entries in the respective matrix power $\bm{A}^p\in\R^{796 \times 796}_{\geq 0}$.
	(a) $\bm{A}, \rho=0.005$, (b) $\bm{A}^2, \rho=0.017$, (c) $\bm{A}^3, \rho=0.05$, (d) $\bm{A}^4, \rho=0.11$, (e) $\bm{A}^5, \rho=0.2$, (f) $\bm{A}^6, \rho=0.317$, (g) $\bm{A}^7, \rho=0.458$, (h) $\bm{A}^8, \rho=0.615$, (i) $\bm{A}^9, \rho=0.767$.}\label{im:sy_powers_multi}
\end{figure}

Fig.~\ref{im:sy_powers_multi} illustrates the sparsity structure of the matrix powers $\bm{A}, \dots , \bm{A}^{9}$ of the undirected layer-coupled Scotland Yard multiplex network with $n=199$ nodes and $L=4$ layers, cf.~Sec.~\ref{sec:Numerical_experiments_approximation_error} for details, and all-to-all inter-layer coupling without self-edges, i.e., $\bm{\tilde{A}}=\bm{1}\bm{1}^T-\bm{I} \in \R^{4 \times 4}$.
Note that layer four in the bottom right block corresponds to the single layer from Fig.~\ref{im:sy_powers_single} but the relative number of non-zeros in the matrix powers $\bm{A}^p$ increases more rapidly than in the single-layer case due to the high degree of connectivity in the inter-layer coupling although all three additional layers are very sparse and non-connected.

Applying the matrix function-based centrality measures from Tab.~\ref{tab:categories_measures} to the supra-adjacency matrix from Eq.~\eqref{eq:supra_adjacency} yields centrality values for all node-layer pairs of the layer-coupled multiplex network.
This allows us to rank the node-layer pairs in terms of their centrality in the network and to identify the most central node-layer pairs.
Following \cite{taylor2017eigenvector} we call the resulting centrality value of the node-layer pair $\xl_i$ the \emph{joint centrality} $JC(i,l)$.
Furthermore, we define \emph{marginal centralities} \cite{taylor2017eigenvector}.
\emph{Marginal node centrality} $MNC(i)$ denotes the importance of a physical node $x_i$ by summing up the joint centralities of its instances across all layers, i.e.,
\begin{equation}\label{eq:MNC}
	MNC(i) = \sum_{l=1}^L JC(i,l).
\end{equation}
Similarly, \emph{marginal layer centrality} $MLC(l)$ denotes the importance of layer $l$ by summing up the joint centralities  of all nodes in this layer, i.e.,
\begin{equation}\label{eq:MLC}
	MLC(l) = \sum_{i=1}^n JC(i,l).
\end{equation}
We illustrate in Sec.~\ref{sec:Numerical_experiments_small_synthetic_network} that aggregation of the layers and subsequent application of matrix function-based centrality measures discards important structural information of the network, which is preserved by the application of multiplex matrix function-based centralities and subsequent summation via marginal centralities.

\section{Efficient methods for computing multiplex centrality measures}\label{sec:Numerical_methods}

Powerful numerical algorithms for the explicit evaluation of functions of small matrices are available in the literature, cf.~e.g., \cite{moler2003nineteen} for the matrix exponential.
These methods, however, become computationally infeasible for medium to large-scale networks.
For these problems, highly efficient methods for approximating certain matrix function expressions have been developed in numerical linear algebra.
In this section, we summarize mostly existing methods based on the approximation of matrix functions by matrix polynomials.
All presented methods utilize the Krylov subspace
\begin{equation}\label{eq:Krylov_subspace}
	\mathcal{K}_k(\bm{A}, \bm{v}) = \text{span}\{ \bm{v} , \bm{A} \bm{v} , \bm{A}^2 \bm{v} , \dots , \bm{A}^{k-1} \bm{v} \},
\end{equation}
which is an intuitive approach to approximating the quantities from Tab.~\ref{tab:categories_measures} in view of Eqs.~\eqref{eq:matrix_exponential} and \eqref{eq:matrix_resolvent}.

The computational bottleneck of all numerical methods presented in this section are matrix-vector products with the matrix $\bm{A}\in\R^{nL \times nL}$ \cite{golub2013matrix}.
Due to the typically encountered sparsity in complex networks, the computational complexity of these matrix-vector products can be assumed to be $\mathcal{O}(nL)$.
If this sparsity assumption is not fulfilled, a linear complexity can still be accomplished if the matrix $\bm{A}$ possesses a low rank factorization or if its entries are determined by certain kernel functions \cite{bergermann2021semi}.
Dense supra-adjacency matrices without exploitable structure, in turn, lead to a complexity of $\mathcal{O}(n^2L^2)$ for each matrix-vector product.

\subsection{Centrality measures based on the evaluation of \texorpdfstring{$f(\bm{A})\bm{b}$}{f(A)b}}\label{sec:Numerical_methods_fAb}

In this section, we summarize established techniques for the approximation of $f(\bm{A})\bm{b}$ with $\bm{A}\in\R^{nL \times nL}$, $\bm{b}\in\R^{nL}$ and where the scalar function $f$ is defined on the spectrum of $\bm{A}$.
These techniques allow the efficient and accurate approximation of total communicability and Katz centrality.

The idea relies on simplifying the problem of computing $f(\bm{A})\bm{b}$ to applying the matrix function $f$ to a reduced matrix $\bm{H}_k\in\R^{k \times k}$ with $k \ll nL$.
It can be shown that if we find an orthogonal matrix $\bm{Q}_k\in\R^{nL \times k}$ such that $\bm{A} \approx \bm{Q}_k \bm{H}_k \bm{Q}_k^T$ we obtain the reduced problem \cite{higham2008functions}
\begin{equation}\label{eq:fAb_arnoldi}
	f(\bm{A})\bm{b} \approx \bm{Q}_k f(\bm{H}_k) \bm{Q}_k^T\bm{b}.
\end{equation}
The columns of the matrix $\bm{Q}_k$ can be obtained by the Arnoldi method \cite{arnoldi1951principle,golub2013matrix} for any real square matrix $\bm{A}$ by iteratively constructing basis vectors of the Krylov subspace defined in Eq.~\eqref{eq:Krylov_subspace} by repeated multiplication of $\bm{A}$ with an iteration vector as well as orthogonalization against the previously computed basis vectors.
The obtained small matrix $\bm{H}_k\in\R^{k \times k}$ has Hessenberg form \cite{golub2013matrix} and the quantity $f(\bm{H}_k)\in\R^{k \times k}$ can cheaply be computed explicitly by standard methods.

In the special case $\bm{A}^T = \bm{A}$, i.e., undirected multiplex networks the basis $\bm{Q}_k$ can similarly be constructed by the Lanczos method \cite{lanczos1950iteration,golub2013matrix}, which reduces $\bm{A}$ to $\bm{A} \approx \bm{Q}_k \bm{T}_k \bm{Q}_k^T$, where $\bm{T}_k\in\R^{k \times k}$ has tridiagonal form.
Any real symmetric matrix $\bm{T}_k^T = \bm{T}_k\in\R^{k \times k}$ has a real-valued eigendecomposition $\bm{T}_k=\bm{S}_k \bm{\Theta}_k \bm{S}_k^T$ with $\bm{S}_k \in \R^{k \times k}$ containing the orthonormal eigenvectors and the diagonal matrix $\bm{\Theta}_k \in \R^{k \times k}$ the eigenvalues of $\bm{T}_k$ as diagonal entries.
In this case, $f(\bm{A})\bm{b}$ can be approximated via
\begin{equation}\label{eq:fAb_lanczos}
	f(\bm{A})\bm{b} \approx \bm{Q}_k \bm{S}_k f(\bm{\Theta}_k) \bm{S}_k^T \bm{Q}_k^T\bm{b},
\end{equation}
where $f(\bm{\Theta}_k)$ applies $f$ elementwise to the eigenvalues of $\bm{T}_k$ \cite{golub2013matrix,higham2008functions}.

The typically encountered linear computational complexity of matrix-vector products with the matrix $\bm{A}\in\R^{nL \times nL}$ described at the beginning of Sec.~\ref{sec:Numerical_methods} makes total communicability and Katz centrality computable to high precision in a matter of seconds even for networks with order $10^7$ node-layer pairs and the limiting factor of the maximally computable network size typically becomes the memory required to store the dense matrix $\bm{Q}_k$.

For the construction of $\bm{Q}_k$ as well as the implementation of Eqs.~\eqref{eq:fAb_arnoldi} and \eqref{eq:fAb_lanczos} we rely on the funm\_kryl toolbox \cite{guttelfunm}, which supports restarted Lanczos and Arnoldi methods \cite{afanasjew2008implementation}.

\subsection{Centrality measures based on the evaluation of \texorpdfstring{$\bm{u}^T f(\bm{A}) \bm{u}$}{uTf(A)u} and \texorpdfstring{$\bm{u}^T f(\bm{A}) \bm{v}$}{uTf(A)v}}\label{sec:Numerical_methods_uTfAu}

This section summarizes mostly existing techniques for the approximation or the computation of lower and upper bounds on subgraph and resolvent-based subgraph centrality, communicability, the Estrada index, and total network communicability, cf.~Tab.~\ref{tab:categories_measures}.
In the symmetric case $\bm{A}^T=\bm{A}$, i.e., undirected multiplex networks we use a well-known relation between Gauss quadrature, the symmetric Lanczos method, and orthogonal polynomials discussed by Golub and Meurant \cite{golub1969calculation,golub1994matrices,golub1997matrices,golub2009matrices}.
In the nonsymmetric case $\bm{A}^T\neq\bm{A}$ we encounter numerical stability issues of standard methods.
We describe two existing approaches to circumvent these issues and propose a third approach, which stabilizes the Arnoldi method by means of a dense shift vector.

While we argued that quantities $f(\bm{A})\bm{b}$ can be computed in $\mathcal{O}(nL)$, in this section, this is only true for total network communicability.
As we have to employ a separate $\mathcal{O}(nL)$ algorithm to approximate each matrix function entry the computational complexity of subgraph, resolvent-based subgraph centrality and the Estrada index is $\mathcal{O}(n^2 L^2)$ and even $\mathcal{O}(n^3 L^3)$ for the computation of all communicabilities.
To circumvent this issue for the trace and the diagonal of $f(\bm{A})$ we present alternative estimation techniques in Sec.~\ref{sec:Numerical_methods_diagonal_trace_estimation}, which are typically faster but less accurate.

\subsubsection{The symmetric case}\label{sec:Numerical_methods_uTfAu_sym}

In order to find lower and upper bounds on quantities of the form $\bm{u}^T f(\bm{A}) \bm{u}$ for $\bm{A}^T=\bm{A} \in \R^{nL \times nL}$, $\bm{u} \in \R^{nL}$, and $f$ a smooth (possibly $C^{\infty}$) function on a given interval on the real line we follow the \emph{Gauss quadrature} approach by Golub and Meurant \cite{golub1969calculation,golub1994matrices,golub1997matrices,golub2009matrices} and consider
\begin{equation}\label{eq:quadrature_eigenvalues}
	\bm{u}^T f(\bm{A}) \bm{u} = \underbrace{\bm{u}^T \bm{\Phi}}_{=:\bm{p}^T} f(\bm{\Lambda}) \underbrace{\bm{\Phi}^T \bm{u}}_{=:\bm{p}} = \bm{p}^T f(\bm{\Lambda}) \bm{p} = \sum_{i=1}^{nL} f(\lambda_i) \bm{p}_i^2,
\end{equation}
for $\bm{A}=\bm{\Phi} \bm{\Lambda} \bm{\Phi}^T$ where $\bm{\Phi} \in \R^{nL \times nL}$ contains the eigenvectors and $\bm{\Lambda}=\text{diag}[\lambda_1, \lambda_2, \dots , \lambda_{nL}]$ the eigenvalues $\lmin=\lambda_1 \leq \lambda_2 \leq \dots \leq \lambda_{nL}=\lmax$ of $\bm{A}$.
Note that this eigendecomposition always exists for symmetric $\bm{A}$.
Furthermore, Eq.~\eqref{eq:quadrature_eigenvalues} can be written as the Riemann-Stieltjes integral
\begin{align*}
	\bm{u}^T f(\bm{A}) \bm{u} & = \int_{\lmin}^{\lmax} f(\lambda)~d\mu(\lambda)\\
	 & = \sum_{j=1}^k w_j f(t_j) + \sum_{m=1}^M v_m f(z_m) + R[f],
\end{align*}
with the remainder $R[f]$, and with the weights $\{ w_j \}_{j=1}^k$ at the $k$ Gauss nodes $\{ t_j \}_{j=1}^k$, the weights $\{ v_m \}_{m=1}^M, M\in\{0,1,2\}$ at the prescribed (interval boundary) nodes $\{ z_m \}_{m=1}^M \subseteq \{\lmin, \lmax\}$, as well as the measure
\begin{equation*}
	\mu(\lambda) = 
	\begin{cases}
		0, & \lambda < \lmin = \lambda_1,\\
		\sum_{j=1}^i \bm{p}_j^2, & \lambda_i \leq \lambda < \lambda_{i+1},\\
		\sum_{j=1}^{nL} \bm{p}_j^2, & \lmax = \lambda_{nL} \leq \lambda.
	\end{cases}
\end{equation*}

Now, using a beautiful relation between Gauss quadrature and orthogonal polynomials constructed from three-term recurrence relations \cite{golub1969calculation,golub1994matrices,golub1997matrices,golub2009matrices} the Gauss nodes and weights do not have to be computed explicitly but can be obtained from a tridiagonalization of the matrix $\bm{A}$, which can be constructed using the Lanczos process, cf.~\cite{lanczos1950iteration}, \cite[Sec.~10.1]{golub2013matrix}, and Sec.~\ref{sec:Numerical_methods_fAb}.
Given the tridiagonal matrix $\bm{T}_k$ after $k$ Lanczos steps it can be shown \cite[Thm.~3.4]{golub1994matrices} that
\begin{equation*}
	\sum_{j=1}^k w_j f(t_j)= \bm{e}_1^T f(\bm{T}_k) \bm{e}_1,
\end{equation*}
with the unit vector $\bm{e}_1 \in \R^k$ where the eigendecomposition $\bm{T}_k=\bm{S}_k \bm{\Theta}_k \bm{S}_k^T$ can be computed cheaply.
We can then easily evaluate $f(\bm{T}_k)=\bm{S}_k f(\bm{\Theta}_k) \bm{S}_k^T$ by elementwise application of $f$ to the eigenvalues $\bm{\Theta}_k$.
Alternatively, the Gauss nodes $t_j$ are given by the eigenvalues of $\bm{T}_k$ in $\bm{\Theta}_k$ and the Gauss weights $w_j$ are given by the squares of the first entries of the respective eigenvectors in $\bm{S}_k$.
This relation yields Gauss quadrature rules, which corresponds to $M=0$.

Additionally, we obtain \emph{Gauss--Radau} ($M=1$) and \emph{Gauss--Lobatto} rules ($M=2$) by prescribing the nodes $z_m$ as eigenvalues to the matrix $\bm{T}_k$, cf.\ \cite[Sec.~3.1]{golub1994matrices} or \cite[Sec.~6.2]{golub2009matrices} for the details.
As the exponential function $f(x)=e^{\beta x}$ and the resolvent function $f(x)=\frac{1}{1-\alpha x}$ with $0 < \alpha < 1/\lmax$ as well as all their derivatives are strictly positive for $x \in [\lmin, \lmax]$ we can determine the sign of the remainder $R[f]$ for all quadrature rules.
More specifically, we obtain lower bounds on $\bm{u}^T f(\bm{A}) \bm{u}$ by the Gauss rule and the Gauss--Radau rule with $z_m=\lmin$ as well as upper bounds by the Gauss--Radau rule with $z_m=\lmax$ and the Gauss--Lobatto rule.

The case $\bm{u}^T f(\bm{A}) \bm{v}$ with $\bm{u} \neq \bm{v}$ can be handled for $\bm{A}^T=\bm{A}$ using the above methods together with the polarization identity \cite{golub2009matrices}
\begin{align}\label{eq:polarization_identity}
	\bm{u}^T f(\bm{A}) \bm{v} = & \frac{1}{4} \left[ (\bm{u} + \bm{v})^T f(\bm{A}) (\bm{u} + \bm{v}) \right. \notag \\
	& \left. \quad - (\bm{u} - \bm{v})^T f(\bm{A}) (\bm{u} - \bm{v}) \right],
\end{align}
at the cost of the evaluation of two quantities of the form $\bm{u}^T f(\bm{A}) \bm{u}$.

\subsubsection{The nonsymmetric case}\label{sec:Numerical_methods_uTfAu_nonsym}

As $\bm{A}^T \neq \bm{A}$ is no longer guaranteed to be diagonalizable Eq.~\eqref{eq:quadrature_eigenvalues} no longer holds in general.
Instead, alternative approaches exist, which are based on the construction of bi-orthogonal polynomials, e.g., by the \emph{nonsymmetric Lanczos method} \cite[Sec.~6.5]{golub2009matrices} or by the \emph{biconjugate gradient method} (BiCG) \cite{saylor2001gaussian} for Gauss quadrature in the complex plane.
Unfortunately, among the quantities $\bm{u}^T f(\bm{A}) \bm{u}$ considered in this section only total network communicability can reliably be computed by these methods.
For the remaining quantities, due to the typically encountered sparsity of the supra-adjacency matrix in combination with the sparsity of the unit vectors as right vectors, we experience serious numerical stability issues with the above methods as well as the Arnoldi method applied only to $f(\bm{A})\bm{u}$.

The described numerical stability issue has previously been discussed in \cite{fenu2013block}.
The author's solution is to first employ the \emph{nonsymmetric block Lanczos method} with an additional dense column in the block vector and then use Gauss and \emph{anti-Gauss} quadrature rules to obtain bounds on the desired quantities.
This \emph{first approach} can be used to compute bounds on quantities of the form $\bm{e}_i^T f(\bm{A}) \bm{e}_i$ in the nonsymmetric case and we refer to \cite{fenu2013block} for the details.
A pleasant side-effect of this method is that we also obtain certain off-diagonal matrix function entries, i.e., communicabilities for free.
However, this approach fails to differentiate between each node-layer pair's role as broadcaster and receiver.

As described in Sec.~\ref{sec:Matrix_centralities} a \emph{second approach} considers the bipartite representation, which exists for any directed network \cite{benzi2020matrix,benzi2013ranking} and which is given by the block supra-adjacency matrix
	\begin{equation}\label{eq:bipartite_adjacency_multi}
	\bm{\mathcal{A}}=
	\begin{bmatrix}
	\bm{0} & \bm{A}\\\bm{A}^T & \bm{0}
	\end{bmatrix} \in \R^{2nL \times 2nL}.
\end{equation}
Obviously, we have $\bm{\mathcal{A}}^T=\bm{\mathcal{A}}$ and thus we can employ the quadrature rules from Sec.~\ref{sec:Numerical_methods_uTfAu_sym} to compute bounds on quantities of the form $\bm{u}^T f(\bm{\mathcal{A}}) \bm{u}$ and in particular $\bm{e}_i^T f(\bm{\mathcal{A}}) \bm{e}_i$ with $\bm{e}_i \in \R^{2nL}$.
Note that there are no stability issues in the symmetric case.
The advantage of this approach is that we obtain broadcaster centralities as the first $nL$ entries of $\bm{e}_i^T f(\bm{\mathcal{A}}) \bm{e}_i$ and receiver centralities as the last $nL$ entries of $\bm{e}_i^T f(\bm{\mathcal{A}}) \bm{e}_i$.

As a \emph{third approach} we propose to utilize the Arnoldi method to approximate $\bm{y}:=f(\bm{A}) \bm{u}$ by adding and subtracting a stabilizing dense shift vector, which ensures numerical stability.
We then obtain the quantity of interest by the additional inner product $\bm{u}^T \bm{y}=\bm{u}^T f(\bm{A}) \bm{u}$.
We choose the one vector $\bm{1}\in\R^{nL}$ as shift vector such that
\begin{equation}\label{eq:shifted_subgraph}
	\bm{e}_i^T f(\bm{A}) \bm{e}_i = \bm{e}_i^T f(\bm{A}) (\bm{e}_i + \bm{1}) - \bm{e}_i^T f(\bm{A}) \bm{1}.
\end{equation}
This choice has the advantage that the quantity $\bm{e}_i^T f(\bm{A}) \bm{1}$ corresponds to total communicability in the case of the matrix exponential and to Katz centrality in the case of the matrix resolvent function, which can be evaluated by one application of the Arnoldi method, cf.~Sec.~\ref{sec:Numerical_methods_fAb}.
We thus obtain an approximation of $\text{diag}(f(\bm{A}))$ at the cost of applying $nL+1$ stable Arnoldi procedures.
Furthermore, we obtain all off-diagonal entries, e.g., communicabilities in the case of the matrix exponential for free.
This approach also fails to differentiate between each node-layer pair's role as broadcaster and receiver, but numerical experiments reveal an interesting error cancellation property:
the approximation error of Eq.~\eqref{eq:shifted_subgraph} tends to be smaller than that of $f(\bm{A}) \bm{1}$ as the subtraction of the two very similar quantities $f(\bm{A}) (\bm{e}_i + \bm{1})$ and $f(\bm{A}) \bm{1}$ cancels part of the approximation error.
This behavior is illustrated in the numerical experiments in Sec.~\ref{sec:Numerical_experiments_approximation_error}.

\subsection{Estimation of the trace and diagonal of \texorpdfstring{$f(\bm{A})$}{f(A)}}\label{sec:Numerical_methods_diagonal_trace_estimation}

When computing bounds on diagonal entries $[f(\bm{A})]_{ii}$ for $i=1, \dots , nL$ and the trace $\text{tr}(f(\bm{A})) = \sum_{i=1}^{nL} [f(\bm{A})]_{ii}$ the $\mathcal{O}(nL)$ computations for each of the $nL$ node-layer pairs result in a total computational complexity of $\mathcal{O}(n^2 L^2)$, which becomes infeasible for medium to large-scale networks.
The authors of \cite{fenu2013network} propose an approach for the efficient identification of the largest diagonal entries of $f(\bm{A})$ for symmetric $\bm{A}$ by means of a low-rank approximation of the matrix $\bm{A}$.
We describe alternative approaches, which rely on the stochastic or deterministic estimation of the trace and the diagonal of matrix functions and which are typically discussed for symmetric $\bm{A}$ in the literature \cite{hutchinson1989stochastic,bekas2007estimator,staar2016stochastic,ubaru2017fast,cortinovis2021randomized,meyer2021hutch++}.
For nonsymmetric $\bm{A}$, i.e., multiplex networks with at least one directed edge we propose to apply the same techniques to the symmetric bipartite network representation defined in Eq.~\eqref{eq:bipartite_adjacency_multi}, which allows to obtain each node-layer pair's broadcaster and receiver centrality.

The classical stochastic \emph{Hutchinson trace estimator} \cite{hutchinson1989stochastic} uses the relation
\begin{equation}\label{eq:hutchinson_estimator}
	\text{tr}(f(\bm{A})) = \sum_{i=1}^{nL} [f(\bm{A})]_{ii} \approx \frac{1}{s} \sum_{k=1}^s \bm{v}_k^T f(\bm{A}) \bm{v}_k,
\end{equation}
where the $\bm{v}_k\in\R^{nL}$ are random \emph{Rademacher vectors} containing the entries $+1$ and $-1$ with probability $1/2$ each.
The goal is to obtain good approximations to $\text{tr}(f(\bm{A}))$ for $s \ll nL$.
Note that the entries of $\bm{v}_k$ can alternatively be sampled from any sub-Gaussian distribution, cf.~\cite{meyer2021hutch++} and references therein.
For symmetric $\bm{A}$ the accuracy of the approximation of $\bm{v}_k^T f(\bm{A}) \bm{v}_k$ has been improved by utilizing the Gauss quadrature techniques introduced in Sec.~\ref{sec:Numerical_methods_uTfAu_sym} for the positive definite \cite{ubaru2017fast} and the indefinite case \cite{cortinovis2021randomized}.
Recently, an improved version of Hutchinson's estimator has been proposed by the inclusion of a low-rank approximation step \cite{meyer2021hutch++}.
Probabilistic error bounds for the respective methods can be found in \cite{hutchinson1989stochastic,ubaru2017fast,cortinovis2021randomized,meyer2021hutch++}.

The authors of \cite{bekas2007estimator} show that Hutchinson's estimator can be extended to approximate the individual diagonal entries of the matrix $f(\bm{A})$ by considering
\begin{equation}\label{eq:diagonal_estimator}
	\text{diag}(f(\bm{A})) \approx \frac{1}{s} \sum_{k=1}^s \bm{v}_k \odot f(\bm{A}) \bm{v}_k,
\end{equation}
where $\odot$ denotes the elementwise vector product and where the factor $\frac{1}{s}$ replaces the elementwise division from \cite[Eq.~(2)]{bekas2007estimator} in the case of vectors $\bm{v}_k$ with entries $\pm 1$.

As Eq.~\eqref{eq:diagonal_estimator} tends to converge slowly \cite{bekas2007estimator}, an alternative (deterministic) approach considers the expression
\begin{equation}\label{eq:diagonal_estimator_hadamard}
	\text{diag}(f(\bm{A})) \approx \frac{1}{s} \left( f(\bm{A}) \odot \bm{V}\bm{V}^T \right) \bm{1},
\end{equation}
which is equivalent to Eq.~\eqref{eq:diagonal_estimator} and where the columns of the matrix $\bm{V}\in\R^{nL \times s}$ are constructed by repeated orthogonal columns of \emph{Hadamard matrices} \cite{bekas2007estimator}.
We refer to the columns of $\bm{V}$ as \emph{Hadamard vectors}.
In this case, the matrix $\bm{V}\bm{V}^T\in\R^{nL \times nL}$ consists of zeros except for the main diagonal and bands with a distance of a multiple of $s$ to the main diagonal.
Hence, Eq.~\eqref{eq:diagonal_estimator_hadamard} computes the row sums of selected bands of $f(\bm{A})$, which converges to $\text{diag}(f(\bm{A}))$ as $s\rightarrow nL$.
However, this approach yields good approximations or even exact results of the diagonal of diagonal-dominant or banded matrices for $s \ll nL$.
The numerical experiments in Sec.~\ref{sec:Numerical_experiments_IMDb} show that the top-ranked node-layer pairs, i.e., largest diagonal entries can be identified even if approximation errors of the numerical values are present.
The computational complexity of $\mathcal{O}(snL)$ resulting from approximating $f(\bm{A}) \bm{v}_k$ from Eq.~\eqref{eq:diagonal_estimator} with the Arnoldi method, cf.~Sec.~\ref{sec:Numerical_methods_fAb}, compares favorably to $\mathcal{O}(n^2 L^2)$ for Gauss quadrature applied to each diagonal entry.
Numerical experiments indicate that sufficiently accurate results can be obtained with $s < \sqrt{nL}$.

In the case of $f(\bm{A}) = \exp(\beta \bm{A})$ and $f(\bm{A}) = (\bm{I} - \alpha \bm{A})^{-1}$ applied to $\bm{A}\in\R^{nL \times nL}_{\geq 0}$ defined in Eq.~\eqref{eq:supra_adjacency} we typically encounter a structure of $f(\bm{A})$, which is dominated by entries on the main diagonal as well as on the diagonals of the off-diagonal blocks.
In this situation, we often obtain good approximations to $\text{diag}(f(\bm{A}))$ for $s \ll nL$ when guaranteeing that the non-zero bands of $\bm{V}\bm{V}^T$ do not coincide with the diagonals of the blocks of $\bm{A}$, i.e., we require $s>L$ and that $n$ is not a multiple of $s$.
As $\bm{A}$ has non-negative entries it can be seen from Eqs.~\eqref{eq:matrix_exponential} and \eqref{eq:matrix_resolvent} that the considered $f(\bm{A})$ also have non-negative entries.
Hence, in our case Eqs.~\eqref{eq:diagonal_estimator} and \eqref{eq:diagonal_estimator_hadamard} yields upper bounds on $\text{diag}(f(\bm{A}))$.

Note that the described techniques have already been used for estimating the trace and the diagonal of single-layer networks \cite{meyer2021hutch++,staar2016stochastic}.
This work, however, is the first to employ the concepts in the context of multiplex networks, which, to the best of our knowledge, is true for all methods presented in Sec.~\ref{sec:Numerical_methods}.

\section{Numerical experiments}\label{sec:Numerical_experiments}

We demonstrate that the proposed generalization of matrix function-based centrality measures to layer-coupled multiplex networks produces meaningful results by applying our methods to several synthetic and real-world multiplex networks.
The evaluation of the quality of centrality rankings is generally difficult as typically there is no ground truth available.
We thus start our discussion with a synthetic undirected and a synthetic directed temporal network, which are designed to produce specific rankings and compare our results to existing centrality measures before we study the performance of the presented numerical methods on several real-world networks with up to $30$ million node-layer pairs.

Matlab code implementing all presented numerical experiments will be made publicly available at \url{https://github.com/KBergermann/Multiplex-matrix-function-centralities}.

\subsection{Small synthetic undirected network}\label{sec:Numerical_experiments_small_synthetic_network}

\begin{figure}
	\centering
	\subfloat[]{
	\includegraphics[width=0.4\textwidth]{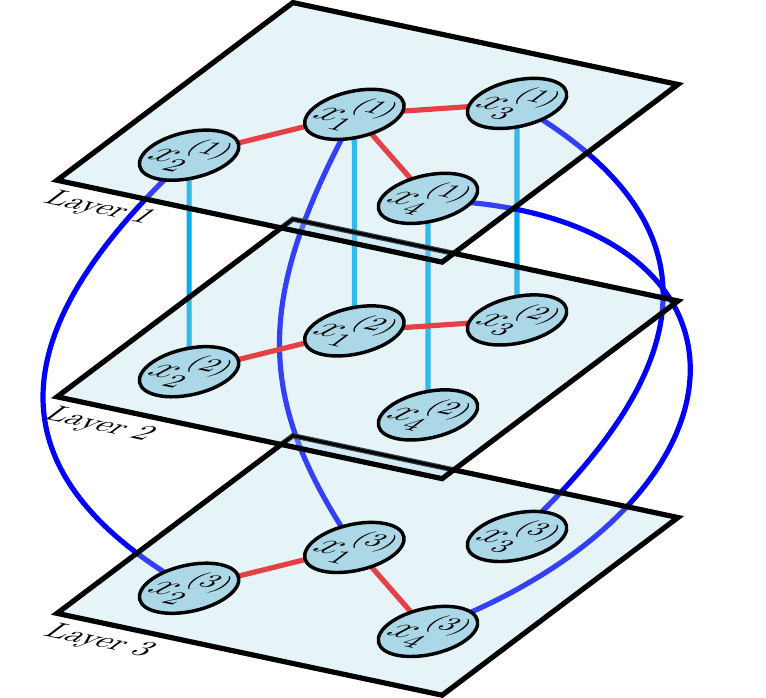}\label{fig:synthetic_network_example_multiplex}
}
\hfill
\subfloat[]{
	\centering
	\includegraphics[width=0.4\textwidth]{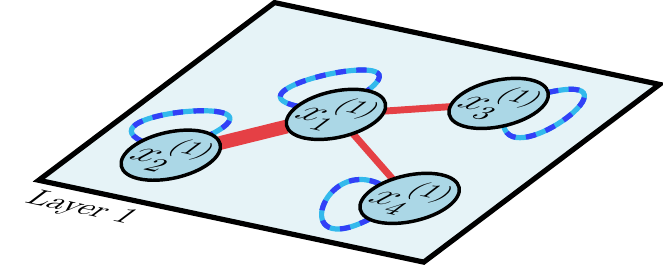}\label{fig:synthetic_network_example_aggregated}
}
	\caption{(a) Example of a layer-coupled multiplex network with $n=4$ physical nodes and $L=3$ layers.
	Intra-layer edges are unweighted and marked in red.
	Inter-layer edges between layers $1$ and $2$ are weighted with $\tilde{\bm{A}}_{12}\geq 0$ and marked in cyan and inter-layer edges between layers $1$ and $3$ are weighted with $\tilde{\bm{A}}_{13}\geq 0$ and marked in blue.
	(b) Aggregated single-layer version of the multiplex network shown in (a).
	Intra-layer edge weights are $3$ between $x_1$ and $x_2$ and $2$ between $x_1$ and $x_3$ as well as $x_1$ and $x_4$.
	In the aggregation, inter-layer edge weights are summed up and added as self-loops with the common value $(\tilde{\bm{A}}_{12}+\tilde{\bm{A}}_{13})$.}\label{fig:synthetic_network_example}
\end{figure}

We start our discussion with the synthetic undirected layer coupled-multiplex network with $n=4$ physical nodes and $L=3$ layers depicted in Fig.~\ref{fig:synthetic_network_example_multiplex}.
The layers consist of unweighted undirected star networks with $x_1$ as center node but with the edge to $x_4$ and $x_3$ removed in layers $2$ and $3$, respectively.
Therefore, by direct comparison of the role of each physical node, the center node $x_1$ can be considered most central followed by $x_2$.
Considering only intra-layer edges, $x_3$ and $x_4$ have the same role in the network giving them an equal centrality.
Furthermore, layers $1$ and $2$ are coupled with undirected edges of weight $\tilde{\bm{A}}_{12}\geq 0$ and layers $1$ and $3$ are coupled with undirected edges of weight $\tilde{\bm{A}}_{13}\geq 0$.
In Fig.~\ref{fig:synthetic_network_example_aggregated} we illustrate the aggregated single-layer version of the multiplex network in which intra-layer edge weights are summed up over the layers and the sum of inter-layer edges is added in the form of self-edges.

\begin{figure}
	\subfloat[]{
	\includegraphics[width=0.48\textwidth]{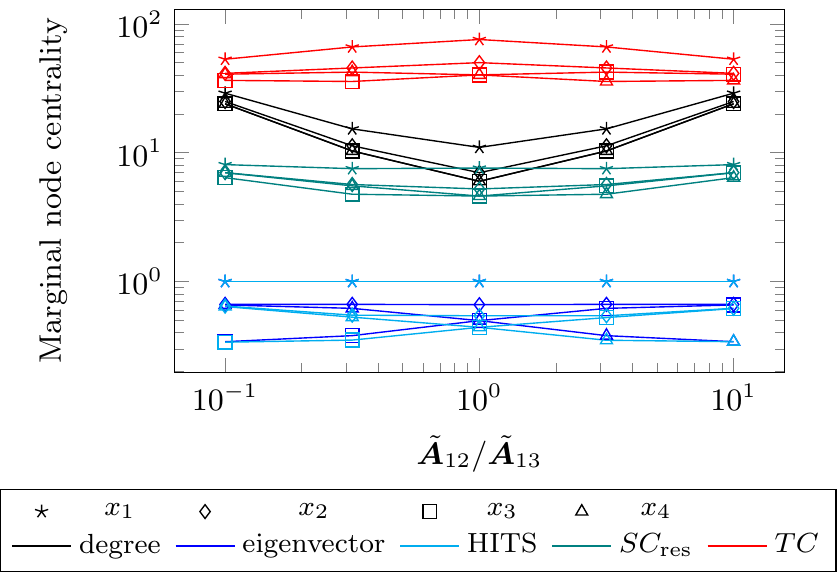}\label{fig:numerics:synthetic_multiplex}
}
\hfill
\subfloat[]{
	\includegraphics[width=0.44\textwidth]{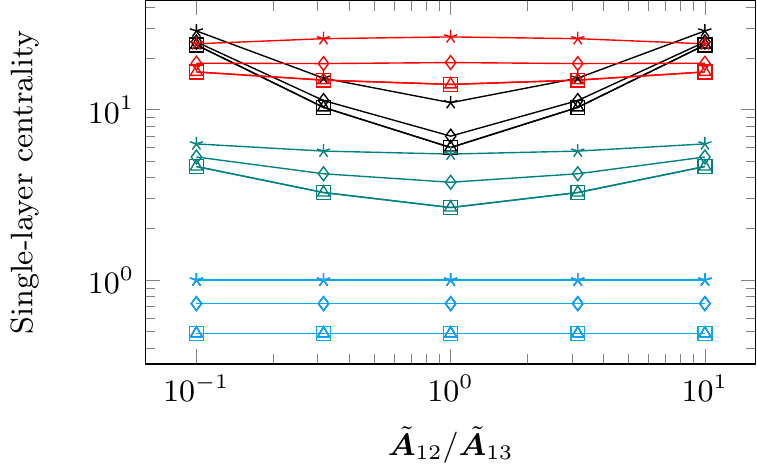}\label{fig:numerics:synthetic_aggregated}
}
	\caption{(a) Marginal node multiplex centralities and (b) single-layer centralities of three established and two matrix function-based centrality measures of the networks depicted in Fig.~\ref{fig:synthetic_network_example}.
	The parameters were chosen $\omega=1$, $\alpha=0.9/\lmax$, and $\beta=3/\lmax$ and for each value of $\tilde{\bm{A}}_{12}/\tilde{\bm{A}}_{13}$ one of the inter-layer weights $\tilde{\bm{A}}_{12}$ and $\tilde{\bm{A}}_{13}$ was fixed at $1$.
	Note that eigenvector centrality and HITS are normalized to have value $1$ for the highest-ranked physical node.}\label{fig:numerics_synthetic}
\end{figure}

We apply total communicability and resolvent-based subgraph centrality to both networks and compare the obtained marginal node centralities with rankings obtained from multilayer versions of degree and eigenvector centrality \cite{de2015ranking,de2015muxviz} as well as HITS (Hyperlink-Induced Topic Search) \cite{kleinberg1999authoritative,de2015muxviz}.
Note that due to the symmetry of $\bm{A}$, HITS returns equal broadcaster and receiver centralities.
The variation of the parameters $\alpha$ and $\beta$ over the intervals $\alpha \in [0.01/\lmax, 0.999/\lmax]$ and $\beta \in [0.01/\lmax, 20/\lmax]$ produced qualitatively coinciding results for the matrix function-based centrality measures.
The results depicted in Fig.~\ref{fig:numerics_synthetic} confirm the presumption that $x_1$ gets ranked first and $x_2$ second by all considered centrality measures.
The ranking of physical nodes $x_3$ and $x_4$ in the multiplex network in Fig.~\ref{fig:numerics:synthetic_multiplex}, however, depends on the ratio $\tilde{\bm{A}}_{12}/\tilde{\bm{A}}_{13}$ of inter-layer weights:
for all measures except degree centrality (here, $x_3$ and $x_4$ have equal centrality independently of $\tilde{\bm{A}}_{12}$ and $\tilde{\bm{A}}_{13}$), $x_3$ and $x_4$ have equal centrality only for $\tilde{\bm{A}}_{12}=\tilde{\bm{A}}_{13}$;
for $\tilde{\bm{A}}_{12}>\tilde{\bm{A}}_{13}$ $x_3$ is more important than $x_4$ and vice versa for $\tilde{\bm{A}}_{12}<\tilde{\bm{A}}_{13}$.

The information about this increased participation in walks around the multiplex network when being connected with an inter-layer edge of high weight in the multiplex case is lost when aggregating the network.
Although all measures in the aggregated network results in Fig.~\ref{fig:numerics:synthetic_aggregated} also rank $x_1$ first and $x_2$ second, $x_3$ and $x_4$ are ranked equally by all measures in the aggregated network.
This example illustrates that multilayer networks are better capable of reflecting complex interactions between its entities than single-layer networks are, that aggregation can discard important structural information, and that the summation of joint centralities to marginal node centralities can not be replaced by the summation of edge weights in a network aggregation process.

\subsection{Small synthetic temporal network}\label{sec:Numerical_experiments_synthetic_temporal_network}

In this subsection, we consider a synthetic temporal network with $n=200$ physical nodes and $L=4$ layers representing different points in time.
For the creation of the network, we follow the procedure described in \cite[Sec.~5.2]{fenu2017block}: intra-layer edges are drawn randomly for all node-layer pairs except for the ``agenda setter'' $x^{(1)}_1$ from which directed paths along four randomly chosen nodes in the four layers distribute information more efficiently than random edges would.

\begin{figure}
	\begin{center}
		\subfloat[]{
			\centering
			\hspace{8pt}
			\includegraphics[width=0.22\textwidth]{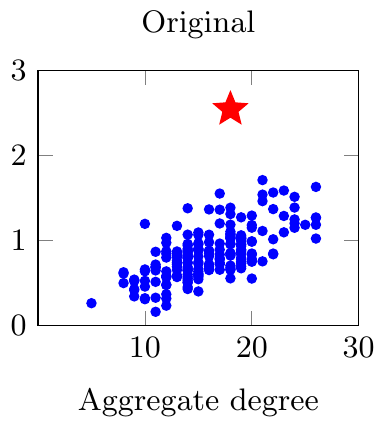}
			\includegraphics[width=0.22\textwidth]{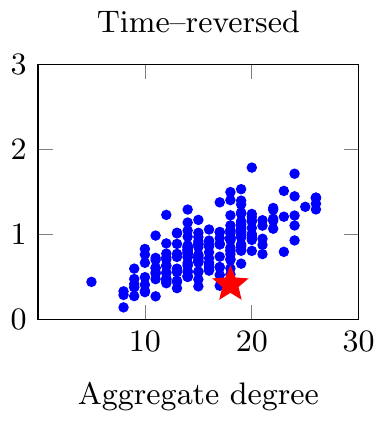}
		}
		\hfill
		\subfloat[]{
			\includegraphics[width=0.24\textwidth]{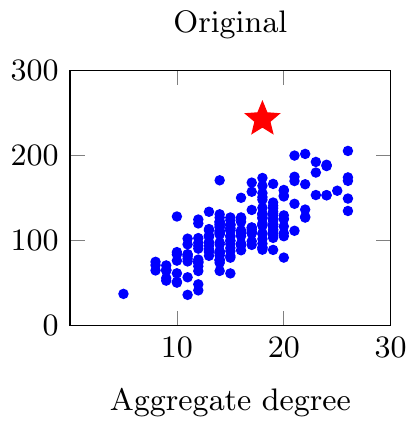}
			\includegraphics[width=0.24\textwidth]{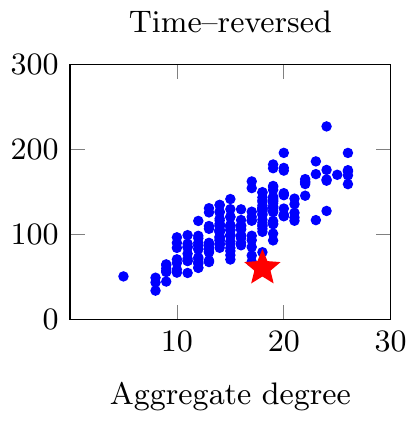}
		}
	\end{center}
	\caption{Comparison of (a) dynamic communicability and (b) marginal node Katz broadcaster centrality using directed temporal coupling with marginal node out-degree centrality of a synthetic temporal network with $n=200$ and $L=4$.
	The red stars represent the ``agenda-setting'' physical node $x_1$, blue dots represent the remaining physical nodes.
	The left plots denote the time layer ordering $1, 2, 3, 4$ and the right plots denote the reversed order.}\label{fig:numerics:synthetic_temporal_network}
\end{figure}

We use the temporal inter-layer coupling described in Sec.~\ref{sec:Multilayer_centralities} and apply Katz centrality with the set of parameters specified in \cite[Sec.~5.2]{fenu2017block}.
This leads to the results depicted in Fig.~\ref{fig:numerics:synthetic_temporal_network} in which dynamic communicability and marginal node Katz broadcaster centrality are plotted against the marginal node out-degree centrality of all physical nodes for forward (Original) and backward (Time--reversed) evolving time.
In this example, the directed temporal coupling in our supra-adjacency matrix framework successfully detects the ``agenda-setting'' property of physical node $x_1$ denoted by the red star in a similar way as dynamic communicability does.

\subsection{Numerical approximation error for small networks} \label{sec:Numerical_experiments_approximation_error}

In order to assess the approximation error of the numerical methods presented in Sec.~\ref{sec:Numerical_methods_fAb} and \ref{sec:Numerical_methods_uTfAu}, we consider one undirected and one directed real-world multiplex network whose sizes still permit the explicit evaluation of the full matrix functions.
We rely on Matlab's \texttt{expm} function and backslash operator for the computation of the ``exact'' matrix function quantities.
The results are, of course, subject to rounding and approximation errors but can be assumed to be highly accurate.

We choose the \emph{Scotland Yard} transportation network created by the authors from a board game as the undirected example network.
It consists of $n=199$ physical nodes representing public transport stops in the city of London and $L=4$ layers representing different modes of transportation (boat, underground, bus, and taxi).
We use all-to-all inter-layer coupling without self-edges.

As directed example network we create a temporal network from the \emph{department 3} subset of the \emph{Email-EU} data set \cite{paranjape2017motifs}.
Each layer represents a time period of $15$ days resulting in $L=35$ layers, which contain the number of Emails exchanged between $n=89$ members of a department of a European research institution as weighted directed edges.
We use directed temporal inter-layer coupling as described in Sec.~\ref{sec:Multilayer_centralities}.

\begin{figure}
	\centering
		\subfloat[]{
		\centering
		\includegraphics[width=0.42\textwidth]{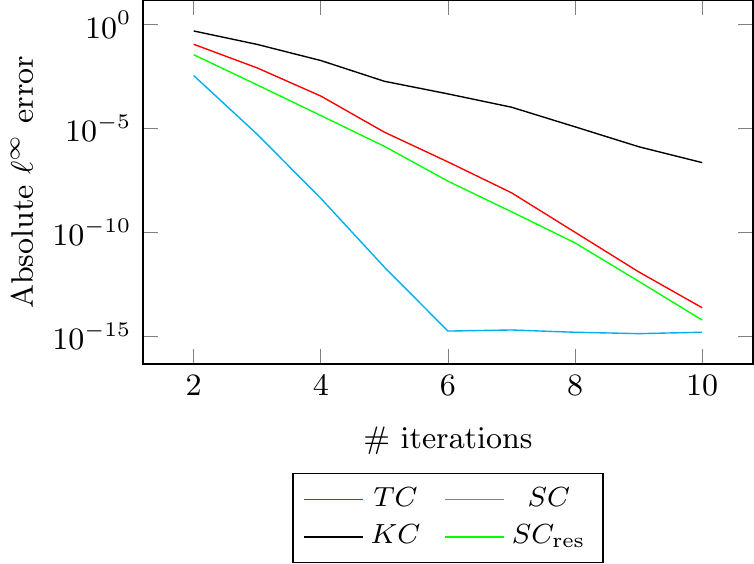}\label{fig:error_plot_symmetric}
	}
\hfill
\subfloat[]{
	\centering
		\includegraphics[width=0.42\textwidth]{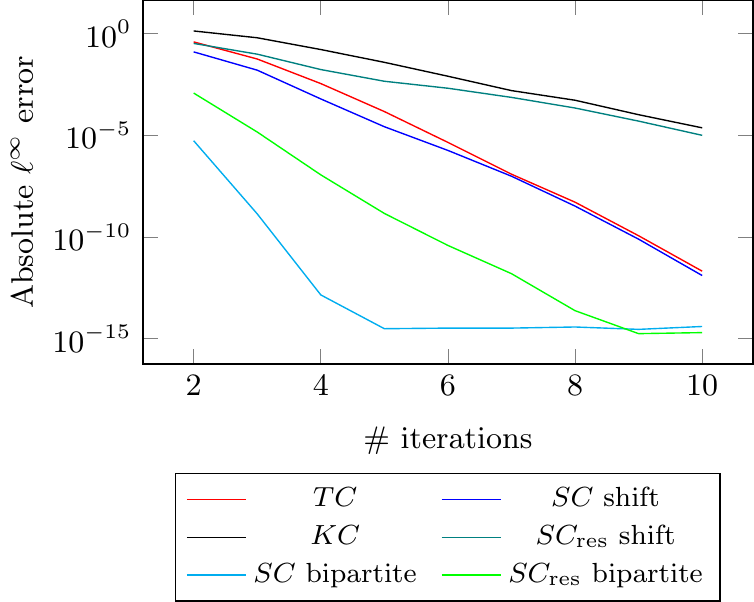}\label{fig:error_plots_nonsymmetric}
}
\caption{Infinity norm error plots of matrix function-based centrality measures of (a) the symmetric supra-adjacency matrix of the Scotland Yard network and (b) the nonsymmetric supra-adjacency matrix of the temporal Email-EU network in dependence of the number of Krylov subspace iterations in the respective methods presented in Secs.~\ref{sec:Numerical_methods_fAb} and \ref{sec:Numerical_methods_uTfAu}.
The parameters are chosen $\omega=1$ and $\alpha=\beta=0.5/\lmax$ for all centrality measures.}\label{fig:error_plots}
\end{figure}

We consider the approximation error
\begin{equation}\label{eq:centrality_error}
	\max_{\substack{i\in \{ 1, \dots ,n\}\\
	l\in \{1, \dots ,L\}}} | XC(i,l,\gamma) - XC_{\mathrm{exact}}(i,l,\gamma) |,
\end{equation}
for $XC\in \{ TC, SC, KC, SC_{\mathrm{res}} \}$ and $\gamma\in \{ \alpha, \beta \}$, which corresponds to the $\ell^{\infty}$ norm error of the vectors of joint centralities.
Fig.~\ref{fig:error_plots} illustrates this error for the two considered networks as a function of the number of iterations of the respective Krylov subspace method.
The plots show that all methods obtain good approximations in only few Krylov subspace iterations.
In general, centrality measures computed by Gauss quadrature rules from Sec.~\ref{sec:Numerical_methods_uTfAu_sym}, e.g., $SC$ and $SC_{\mathrm{res}}$ in Fig.~\ref{fig:error_plot_symmetric} and $SC$ bipartite and $SC_{\mathrm{res}}$ bipartite in Fig.~\ref{fig:error_plots_nonsymmetric} converge faster than the remaining measures relying on the approximation of a quantity $f(\bm{A})\bm{b}$ by a method from Sec.~\ref{sec:Numerical_methods_fAb} or Sec.~\ref{sec:Numerical_methods_uTfAu_nonsym}.
This observation in line with theoretical results that after $k$ Krylov iterations the methods presented in Sec.~\ref{sec:Numerical_methods_fAb} to approximate $f(\bm{A})\bm{b}$ interpolate polynomials of degree $k-1$ exactly while the different Gauss quadrature rules from Sec.~\ref{sec:Numerical_methods_uTfAu_sym} for quantities $\bm{u}^T f(\bm{A}) \bm{u}$ interpolate polynomials of degrees between $2k-1$ and $2k+1$ exactly \cite{golub2009matrices}.

Furthermore, for our choice of $\alpha=\beta$ the quantities based on the matrix exponential converge faster than those based on the matrix resolvent function.
This is due to the factor $\frac{1}{p!}$ in the power series of the matrix exponential, which lets the contribution of high matrix powers decay more rapidly than in the case of the matrix resolvent and thus enables better approximations of $f(\bm{A})$ with low-order polynomials.

Finally, the slightly faster convergence of $SC$ shift compared to $TC$ and $SC_{\mathrm{res}}$ shift compared to $KC$ in Fig.~\ref{fig:error_plots_nonsymmetric} illustrates the error cancellation property described in Sec.~\ref{sec:Numerical_methods_uTfAu_nonsym}:
the subtraction of two similar quantities of the form $f(\bm{A})\bm{b}$ in Eq.~\eqref{eq:shifted_subgraph} annihilates part of the approximation error.

\subsection{Medium-sized European airlines network}\label{sec:Numerical_experiments_European_airlines}

The European airlines data set \cite{cardillo2013emergence} consists of 450 physical nodes representing European airports and $L=37$ layers representing European airlines.
The network is unweighted and undirected, i.e., the symmetric intra-layer adjacency matrices $\A1, \dots ,\bm{A}^{(37)}$ contain ones where the respective airline offers a flight connection between two airports and zeros otherwise.
We choose all-to-all layer coupling without self-edges, i.e., $\tilde{\bm{A}}=\bm{1}\bm{1}^T - \bm{1}$ to reflect the effort added by changing airlines on connecting flights between any pair of distinct airlines.
We only include the $n=417$ airports belonging to the largest connected cluster in the sum of the intra-layer adjacency matrices in order to be able to compare our results with eigenvector centralities from \cite[Sec.~5.1]{taylor2021tunable}.
While this selection is a necessary requirement for the supra-adjacency matrix to satisfy the assumptions of the Perron--Frobenius theorem, which guarantees the unique existence of the largest eigenvector of the matrix \cite[Thm.~3.7]{taylor2021tunable}, this restriction would not be required for our matrix function-based centrality framework.

Very few iterations of the methods described in Sec.~\ref{sec:Numerical_methods_fAb} and \ref{sec:Numerical_methods_uTfAu_sym} already achieve a notable precision.
Tab.~\ref{tab:airlines_estrada_bounds} illustrates lower and upper quadrature bounds on the Estrada index defined in Tab.~\ref{tab:categories_measures} for the parameters $\omega=1$ and $\beta=5/\lmax$, which lead to $\lmax \approx 38.37$.
Note that the Estrada index is a sum of $nL=15\,429$ individually computed quantities of the form $\bm{u}^T f(\bm{A}) \bm{u}$.
The computations require a total of $87$ seconds using $5$ Lanczos iterations per node-layer pair whereas Matlab's \texttt{expm} requires $460$ seconds.
All runtime measurements in this subsection were performed on a laptop with 16 GB RAM and an Intel Core i5-8265U CPU with 4 $\times$ 1.60--3.90 GHz cores and Matlab R2020b.

\begin{table}
	\begin{center}
		\begin{tabular}{ c @{\hspace{6pt}} c @{\hspace{6pt}} c @{\hspace{6pt}} c @{\hspace{6pt}} c @{\hspace{6pt}} c}
			\hline \hline
			\# iterations & 1 & 2 & 3 & 4 & 5\\
			\hline
			G (lower) & 15\,429 & 58\,116 & 58\,761 & 58\,770.66 & 58\,770.9769\\
			GR (lower) & 18\,976 & 58\,632 & 58\,769 & 58\,770.90 & 58\,770.9832\\
			GR (upper) & 68\,641 & 59\,140 & 58\,777 & 58\,771.04 & 58\,770.9846\\
			GL (upper) & 542\,417 & 65\,837 & 58\,865 & 58\,771.91 & 58\,770.9906\\
			\hline \hline
		\end{tabular}
		\caption{Bounds on the multiplex Estrada index in dependence of the number of Lanczos iterations for the unweighted undirected European airlines data set with all-to-all inter-layer coupling without self-edges using Gauss (G), Gauss--Radau (GR) and Gauss--Lobatto (GL) quadrature rules, cf.~Sec.~\ref{sec:Numerical_methods_uTfAu_sym}, with $\omega=1$ and $\beta=5/\lmax$.}\label{tab:airlines_estrada_bounds}
	\end{center}
\end{table}

\begin{figure}
	\includegraphics[width=0.46\textwidth]{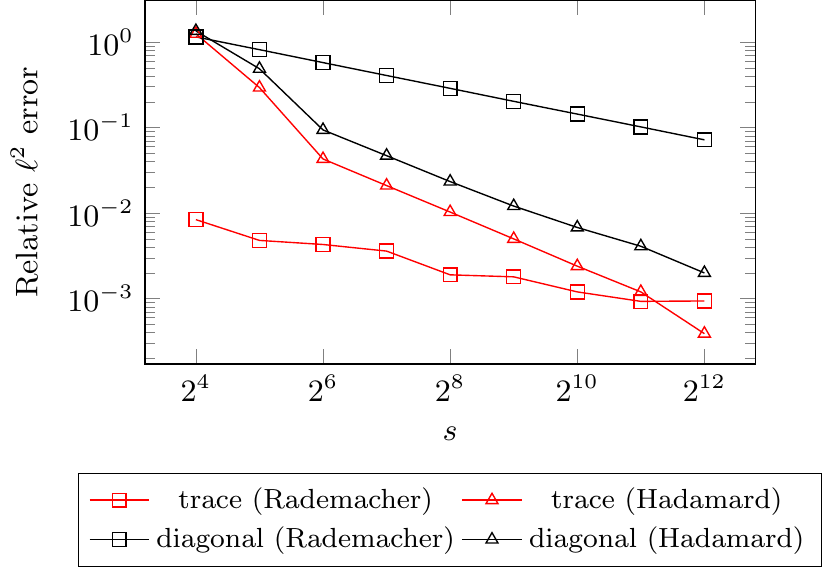}
	\caption{Relative error of the stochastic (Rademacher) and deterministic (Hadamard) estimation of the Estrada index (trace) and subgraph centrality (diagonal) for the European airlines network with $\omega=1$, $\beta=5/\lmax$, and $\tilde{\bm{A}}=\bm{1}\bm{1}^T - \bm{I}$.
	The abscissa $s$ denotes the number of Rademacher and Hadamard vectors, respectively, cf.~Sec.~\ref{sec:Numerical_methods_diagonal_trace_estimation}.
	For the relative error of the diagonal of $f(\bm{A})$ we use the $\ell^2$ norm of the vector of subgraph centralities.
	The results using stochastic Rademacher vectors are averaged over $10$ independent runs.}\label{fig:airlines_trace_diagonal_error}
\end{figure}

Even lower runtimes can be achieved by employing the trace and diagonal estimation techniques described in Sec.~\ref{sec:Numerical_methods_diagonal_trace_estimation}.
Fig.~\ref{fig:airlines_trace_diagonal_error} demonstrates that an approximation of the Estrada index to a relative error below $1\%$ can be obtained using the Hutchinson estimator from Eq.~\eqref{eq:hutchinson_estimator} with only $s=16$ Rademacher vectors.
This requires around $0.4$ seconds runtime when the funm\_kryl toolbox for Matlab \cite{guttelfunm} with $20$ Lanczos iterations is used to evaluate matrix-vector products of the matrix $\bm{A}$ with the Rademacher vectors to a precision of $10^{-12}$.
Increasing $s$ leads to a linear increase in runtime such that the computational cost of the estimation surpasses that of applying Gauss quadrature to each diagonal entry for $s=2^{12}$, which requires around $95$ seconds.
Note that the relative error of the Gauss--Radau bounds from Tab.~\ref{tab:airlines_estrada_bounds} with $5$ Lanczos iterations is of order $10^{-8}$.
Comparing Rademacher and Hadamard vectors, Fig.~\ref{fig:airlines_trace_diagonal_error} shows that while Rademacher vectors are better suited for trace estimation with low $s$, Hadamard vectors perform better in estimating the diagonal of $f(\bm{A})$ if the conditions on $s$ discussed in Sec.~\ref{sec:Numerical_methods_diagonal_trace_estimation} are fulfilled.

\begin{table*}
	\begin{center}
		\begin{tabular}{c @{\hspace{20pt}} c @{\hspace{20pt}} c}
			\hline \hline
			Katz centrality & Eigenvector centrality \cite{taylor2021tunable} & Degree centrality\\ \hline
			\begin{tabular}{lc}
				(Stansted, Ryanair) & $4.4231$\\
				(Munich, Lufthansa) & $4.0939$\\
				(Frankfurt, Lufthansa) & $4.0652$\\
				(Atat\"urk, Turkish) & $4.0488$\\
				(Gatwick, easyJet) & $3.7927$\\
				(Dublin, Ryanair) & $3.6481$\\
				(Vienna, Austrian) & $3.5941$\\
				(Amsterdam, KLM) & $3.5663$\\
				(Bergamo, Ryanair) & $3.3246$\\ 
				(Paris, Air France) & $3.2446$\\ 
			\end{tabular} & 
			\begin{tabular}{lc}
				(Frankfurt, Lufthansa) & $0.0638$\\
				(Munich, Lufthansa) & $0.0631$\\
				(Amsterdam, KLM) & $0.0564$\\
				(D\"usseldorf, Lufthansa) & $0.0530$\\
				(Madrid, Iberia) & $0.0504$\\
				(Vienna, Austrian) & $0.0490$\\
				(Paris, Air France) & $0.0485$\\
				(Madrid, Ryanair) & $0.0482$\\
				(Gatwick, easyJet) & $0.0472$\\
				(Fuimicino, Alitalia) & $0.0471$\\
			\end{tabular} & 
			\begin{tabular}{lc}
				(Stansted, Ryanair) & $121$\\
				(Atat\"urk, Turkish) & $118$\\
				(Munich, Lufthansa) & $114$\\
				(Frankfurt, Lufthansa) & $113$\\
				(Gatwick, easyJet) & $103$\\
				(Vienna, Austrian) & $100$\\
				(Amsterdam, KLM) & $98$\\
				(Dublin, Ryanair) & $90$\\
				(Paris, Air France) & $86$\\
				(Fuimicino, Alitalia) & $84$\\
			\end{tabular}\\ \hline \hline
		\end{tabular}
		\caption{Top 10 joint centralities of the unweighted undirected European airlines data set with all-to-all inter-layer coupling and parameter $\omega=1$.
		The columns contain Katz centrality with the parameter  $\alpha=0.5/\lmax$, eigenvector centrality computed with the codes \cite{taylorsupracentrality}, and degree centrality where each degree includes $L-1=36$ inter-layer edges of weight 1 from inter-layer coupling.}\label{tab:ea_joint}
	\end{center}
\end{table*}

In Tab.~\ref{tab:ea_joint} we display the top $10$ node-layer pairs of Katz centrality computed with parameters $\omega=1$ and $\alpha=0.5/\lmax$ and compare it with eigenvector \cite{taylor2021tunable,taylorsupracentrality} and degree centrality.
The results show that degree centrality ranks the centers of (almost) star network layers like Atat\"urk airport in the Turkish airlines layer highly while eigenvector centrality favors, e.g., Frankfurt and Munich in the Lufthansa layer, which are themselves connected to many other central node-layer pairs.
As the matrix function-based centrality measures interpolate between these two established concepts \cite{benzi2015limiting} and the parameter $\alpha=0.5/\lmax$ is chosen in the middle of its admissible interval, Katz centrality rankings lie in between the results from degree and eigenvector centrality.
Note that subgraph and resolvent-based subgraph centrality as well as total communicability similarly interpolate between eigenvector and degree centrality.

It is interesting to note that the top three Katz marginal node centralities include Madrid and Barcelona airport although both airports do not enter the top joint centrality rankings.
This situation can only emerge in a constellation where these two airport's importance is well-spread over many different layers, i.e., airlines.

\begin{figure}
	\centering
	\subfloat[]{
		\includegraphics[width=0.44\textwidth]{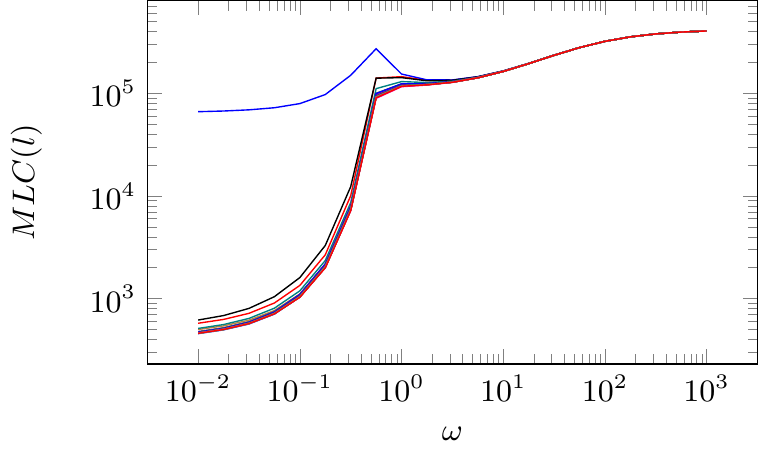}\label{im:ea_KC_MLC}
	}
	\hfill
	\subfloat[]{
		\includegraphics[width=0.44\textwidth]{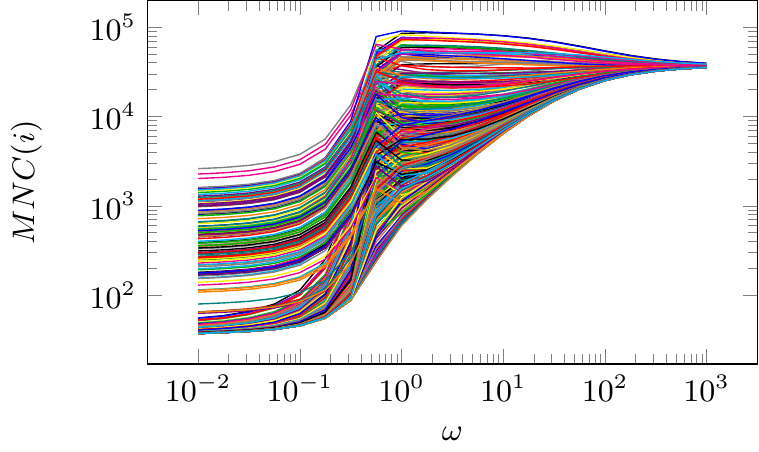}\label{im:ea_KC_MNC}
	}
	\caption{(a) Marginal layer and (b) marginal node Katz centralities for the unweighted undirected European airlines data set with all-to-all inter-layer coupling without self-edges, $\alpha=0.999/\lmax$, varying coupling parameter $\omega$, and $l=1, \dots , L$ and $i=1, \dots , n$, respectively.}\label{im:ea_KC}
\end{figure}

For this network, the obtained marginal node and layer rankings barely depend on the hyper-parameters $\alpha$, $\beta$, and $\omega$.
The only exception from this behavior can be observed in the limit $\alpha \rightarrow \frac{1}{\lmax}^-$ in which resolvent-based subgraph and Katz centrality converge to eigenvector centrality \cite{benzi2015limiting}.
Figs.~\ref{im:ea_KC_MLC} and \ref{im:ea_KC_MNC} show that for $\alpha=0.999/\lmax$ the variation of the parameter $\omega$ in the medium coupling regime, i.e., values around $\omega=1$ leads to a strong reordering in marginal node centralities.
This qualitative behavior, which includes a clustering of marginal node centralities in the weak coupling regime, i.e., $\omega \rightarrow 0^+$ is similar to that observed for marginal node eigenvector centralities of multiplex networks in \cite[Fig.~5]{taylor2021tunable}.
However, an interesting qualitative difference is the convergence towards a common limit value in the strong coupling limit, i.e., $\omega \rightarrow \infty$ in Fig.~\ref{im:ea_KC_MNC} compared to the convergence towards different limit values in \cite[Fig.~5]{taylor2021tunable}.
Marginal layer centralities of matrix function-based and eigenvector centrality measures behave very similarly including the dominance of the Ryanair layer in the weak coupling regime, which is caused by the fact that it contains by far the largest number of intra-layer edges.

The potential runtime gains of trace and diagonal estimation techniques from Sec.~\ref{sec:Numerical_methods_diagonal_trace_estimation} compared to the explicit evaluation of the matrix functions discussed earlier also markedly come to light for the numerical methods from Secs.~\ref{sec:Numerical_methods_fAb} and \ref{sec:Numerical_methods_uTfAu_sym}.
The quantity $f(\bm{A})\bm{b}$ for Katz centrality or total communicability can be approximated within $0.06$ seconds employing $30$ Lanczos iterations using the funm\_kryl toolbox for Matlab \cite{guttelfunm} to a precision of $10^{-16}$.
The $87$ seconds runtime for computing subgraph and resolvent-based subgraph centrality for all $nL=15\,429$ node-layer pairs discussed earlier can be reduced further by employing a very straightforward parallelization using the parallel for-loop \texttt{parfor} from the Matlab Parallel Computing Toolbox.
Here, using $4$ processors the total runtime reduces to $35$ seconds plus a one-time setup of the processes, which requires around $20$ seconds.

\subsection{Large temporal IMDb Sci-Fi network}\label{sec:Numerical_experiments_IMDb}

As the final example network we consider the collaboration network of the principal cast and crew members of all science fiction (Sci-Fi) movies and series episodes from the publicly available Internet Movie Database (IMDb) data set \cite{IMDb}.
A total of $n=245\,757$ principals is involved in $96\,838$ Sci-Fi movies or episodes with release dates between the years $1895$ and $2028$ (some release dates lie in the future as the IMDb contains records of unreleased but announced titles).
We build a temporal network with each layer representing one release year.
We exclude years in which no collaboration took place, e.g., $1895$ contains only one movie with one principal.
This leads to a total of $L=124$ layers and thus $nL=30\,473\,868$ node-layer pairs.
The undirected intra-layer edges are weighted with the number of collaborations between all pairs of principals in the given year.
For Sci-Fi series we count seasons as collaborations as counting episodes leads to an undesired dominance of series with many episodes in the rankings.
For the inter-layer coupling we use directed temporal coupling with the weights $\tilde{\bm{A}}_{(l-1),l} = e^{- \Delta t_l}, l=2, \dots, L$ with $\Delta t_l$ denoting the time difference in years between layers $l-1$ and $l$.

\begin{table}
	\begin{center}
		\begin{tabular}{c @{\hspace{5pt}} cccc @{\hspace{5pt}} cccc}
			\hline \hline
			Principals & \multicolumn{4}{c}{broadcaster} & \multicolumn{4}{c}{receiver} \\ \cline{2-5} \cline{5-9}
			& $TC$ & $KC$ & $SC$ & $SC_{\mathrm{res}}$ & $TC$ & $KC$ & $SC$ & $SC_{\mathrm{res}}$\\ \hline
			Vance Major & $1$ & $1$ & $1$ & $1$ & $1$ & $1$ & $1$ & $1$ \\ 
			Adam Mullen & $2$ & $2$ & $2$ & $2$ & $2$ & $2$ & $2$ & $2$\\ 
			Kevin MacLeod & $15$ & $3$ & $17$ & $17$ & $15$ & $3$ & $16$ & $16$ \\ 
			Gene Roddenberry & $16$ & $4$ & $16$ & $16$ & $16$ & $4$ & $17$ & $17$ \\ 
			George Lucas & $29$ & $19$ & $60$ & $50$ & $29$ & $19$ & $57$ & $48$ \\ 
			William Shatner & $41$ & $29$ & $52$ & $47$ & $41$ & $29$ & $52$ & $45$ \\ 
			Jack Kirby & $38$ & $27$ & $63$ & $42$ & $38$ & $27$ & $65$ & $68$ \\ 
			H.G. Wells & $43$ & $25$ & $66$ & $49$ & $43$ & $25$ & $79$ & $78$ \\ 
			Leonard Nimoy & $99$ & $56$ & $128$ & $134$ & $100$ & $56$ & $108$ & $84$ \\ 
			Jules Verne & $113$ & $67$ & $160$ & $183$ & $114$ & $67$ & $117$ & $60$ \\ 
			Kate Mulgrew & $106$ & $92$ & $102$ & $116$ & $106$ & $92$ & $104$ & $119$ \\ 
			James Cameron & $118$ & $71$ & $147$ & $161$ & $117$ & $71$ & $149$ & $160$ \\ 
			Stephen King & $150$ & $91$ & $235$ & $248$ & $149$ & $91$ & $235$ & $248$ \\ 
			Patrick Stewart & $164$ & $108$ & $257$ & $294$ & $164$ & $108$ & $252$ & $293$ \\ 
			\hline \hline
		\end{tabular}
		\caption{Marginal node centrality ranks of selected physical nodes (principals) of the temporal IMDb Sci-Fi network with $\omega=10$, $\alpha=0.9/\lmax$, and $\beta=5/\lmax$.
		$SC$ and $SC_{\mathrm{res}}$ were computed via diagonal estimation with $s=1024$, cf.~Sec.~\ref{sec:Numerical_methods_diagonal_trace_estimation}.}\label{tab:IMDb_MNCs}
	\end{center}
\end{table}

As the directed inter-layer edges make the supra-adjacency matrix $\bm{A}$ nonsymmetric we separately consider broadcaster and receiver centralities.
However, as all intra-layer edges are undirected there are almost no differences between $TC$ and $KC$ broadcaster and receiver centralities and $SC$ and $SC_{\mathrm{res}}$ rankings also tend to be similar.
Tab.~\ref{tab:IMDb_MNCs} lists marginal node centrality rankings of selected principals for $\omega=10$, which leads to $\lmax \approx 74.6$.
Marginal layer centralities increase with the increasing volume of produced Sci-Fi content over the years but drop substantially in layers in which the time difference to the following layer is more than one year.
The variation of the coupling parameter $\omega$ in the interval $[10^{-2}, 10^2]$ shows a tendency that larger values of $\omega$ favor principals active in early time layers.
Its choice thus controls the knock-on effect, i.e., the influence of early works on later productions.

Albeit the large network size, the approximation of the quantities $TC$ and $KC$ to a precision of $10^{-16}$ in funm\_kryl \cite{guttelfunm} can be achieved in $17.1$ and $46.5$ seconds for $30$ and $80$ Krylov iterations, respectively.
All runtime measurements in this subsection were performed on a computer with 32 GB RAM and an Intel Core i7-4770  with 4 $\times$ 3.4 GHz cores and Matlab R2021b.

\begin{figure}
	\includegraphics[width=0.43\textwidth]{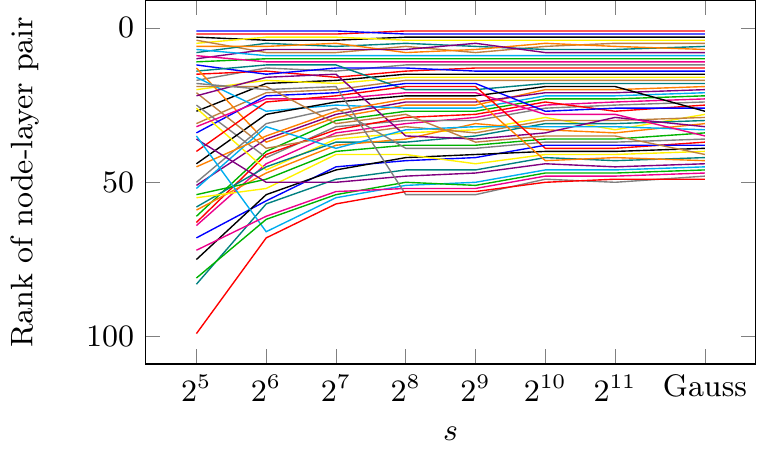}
	\caption{Rankings of the top $50$ joint broadcaster subgraph centralities of the temporal IMDb Sci-Fi network using Hadamard diagonal estimation.
	Numerical values for $s$ denote the employed number of Hadamard vectors, cf.~Sec.~\ref{sec:Numerical_methods_diagonal_trace_estimation}, and ``Gauss'' denotes results computed to high precision with Gauss quadrature rules, cf.~Sec.~\ref{sec:Numerical_methods_uTfAu_sym}.
	The parameters are chosen $\omega=10$ and $\beta=5/\lmax$.}\label{fig:IMDb_diagonal_estimation_rankings}
\end{figure}

\begin{figure}
	\includegraphics[width=0.43\textwidth]{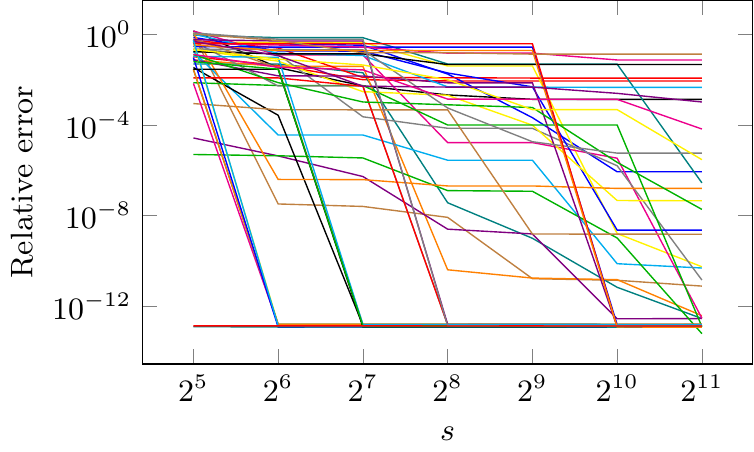}
	\caption{Relative approximation error of the top $50$ joint broadcaster subgraph centralities of the temporal IMDb Sci-Fi network using Hadamard diagonal estimation.
	The true centrality values are computed to high precision with Gauss quadrature, cf.~Sec.~\ref{sec:Numerical_methods_uTfAu_sym}.
	The abscissa $s$ denotes the number of Hadamard vectors, cf.~Sec.~\ref{sec:Numerical_methods_diagonal_trace_estimation}.
	The parameters are chosen $\omega=10$ and $\beta=5/\lmax$.}\label{fig:IMDb_diagonal_estimation_errors}
\end{figure}

The computation of one diagonal entry of $f(\bm{\mathcal{A}})$ for $SC$ and $SC_{\mathrm{res}}$ using $10$ Lanczos iterations requires $14.6$ seconds.
With this approach the sequential computation of the full diagonal of $f(\bm{\mathcal{A}})$ would require approximately $28.2$ years corroborating the infeasibility of this approach for large-scale networks.
Employing the deterministic diagonal estimation from Sec.~\ref{sec:Numerical_methods_diagonal_trace_estimation} leads to $31.8$ seconds runtime per $s$, i.e., per Hadamard vector when the matrix-vector products of $f(\bm{A})$ with the Hadamard vectors are computed to a precision of $10^{-12}$ using funm\_kryl \cite{guttelfunm}.
Figs.~\ref{fig:IMDb_diagonal_estimation_rankings} and \ref{fig:IMDb_diagonal_estimation_errors} show that relatively small numbers $s$ of Hadamard vectors achieve approximations, which reliably identify the top-ranked node-layer pairs even when non-negligible diagonal estimation errors are present.
However, for a portion of the considered node-layer pairs, which increases with increasing $s$, the centrality value is computed to high precision, cf.~Fig.~\ref{fig:IMDb_diagonal_estimation_errors}.
If the memory requirement of storing the matrix $\bm{V}\in\R^{nL \times s}$, cf.~Sec.~\ref{sec:Numerical_methods_diagonal_trace_estimation}, becomes a limiting factor in the computations the columns of $\bm{V}$ can be assembled on-the-fly using Kronecker products of small Hadamard matrices as described in \cite{bekas2007estimator} at little extra cost.
Accurate bounds on a few pre-selected node-layer pairs can be computed relatively cheaply using Gauss quadrature techniques.

\section{Conclusion and Outlook}\label{sec:Conclusion_outlook}

We presented a general framework to apply matrix function-based centrality measures to layer-coupled multiplex networks using the supra-adjacency matrix as network representation.
We employed highly scalable numerical methods, which enable the efficient treatment of medium to large-scale networks.
The application to several synthetic and real-world multiplex networks and the comparison with established multilayer centrality measures indicate that our approach obtains sensible rankings of nodes, layers, and node-layer pairs for weighted and unweighted as well as directed and undirected multiplex networks in competitive runtimes.
The influence of the involved hyper-parameters was discussed in various example scenarios.

Besides the application to complex inherently multilayered problems, future work could center around more general multilayer network models and different network representations, e.g., in the form of tensors.
Moreover, the influence of the special structure of the supra-adjacency matrix on the convergence of the numerical methods could be studied.

\section*{Acknowledgments}

M.~Stoll acknowledges the funding of the BMBF grant 01--S20053A.
The authors thank all referees for their helpful comments.

\end{document}